\documentclass[12pt]{article}
\usepackage{fullpage,hyperref,amssymb,amsmath}

\def\a{\alpha}\def\b{\beta}\def\g{\gamma}\def\d{\delta}\def\e{\epsilon}
\def\k{\kappa}\def\l{\lambda}\def\o{\omega}
\def\r{\rho}\def\s{\sigma}\def\t{\tau}\def\u{\upsilon}\def\z{\zeta}\def\th{\theta}
\def\varphi{\varphi}
\def\G{\Gamma}\def\D{\Delta}\def\L{\Lambda}\def\O{\Omega}

\def\CB{\mathcal{B}}
\def\CE{\mathcal{E}}\def\CF{\mathcal{F}}
\def\CI{\mathcal{I}}
\def\CJ{\mathcal{J}}
\def\CM{\mathcal{M}}\def\CN{\mathcal{N}}

\def\CX{\mathcal{X}}

\def\IZ{\mathbb{Z}}
\def\IR{\mathbb{R}}\def\IC{\mathbb{C}}
\def\IP{\mathbb{P}}

\def\so{\mathfrak{so}}
\def\su{\mathfrak{su}}

\def\jbar{{\bar\jmath}}
\def\kbar{{\bar k}}
\def\mbar{{\bar m}}\def\nbar{{\bar n}}
\def\tbar{{\bar t}}

\def\pd{\partial}

\def\bt{\mathbf{t}}
\def\bv{\mathbf{v}}
\def\bx{\mathbf{x}}

\DeclareMathOperator{\re}{Re}  % \Re defined by AMS LaTeX
\DeclareMathOperator{\im}{Im}  % \Im defined by AMS LaTeX
\DeclareMathOperator{\Spin}{Spin}
\DeclareMathOperator{\Harm}{Harm}

\DeclareMathOperator{\Jac}{Jac}
\DeclareMathOperator{\Pic}{Pic}

\DeclareMathOperator{\Vol}{Vol}
\DeclareMathOperator{\MW}{MW}

\DeclareMathOperator{\GL}{GL}
\DeclareMathOperator{\SL}{SL}
\DeclareMathOperator{\SO}{SO}
\DeclareMathOperator{\U}{U}
\DeclareMathOperator{\SU}{SU}

\DeclareMathOperator{\PSL}{PSL}

\def\tilde{\widetilde}

\def\w{\wedge}

\def\half{\tfrac12}

\def\free{\text{free}}
\def\tor{\text{tor}}

\long\def\symbolfootnote[#1]#2{\begingroup%
\def\thefootnote{\fnsymbol{footnote}}\footnote[#1]{#2}\endgroup}

\def\K3{\text{K3}}
\def\tw{\text{tw}}

\def\Gbar{{\bar G}}
\def\Gtilde{{\tilde G}}
\def\mhat{{\hat m}}
\def\nhat{{\hat n}}
\def\phat{{\hat p}}
\def\yhat{{\hat y}}
\def\ahat{{\hat\a}}
\def\bhat{{\hat\b}}

\def\mhat{{\hat m}}
\def\nhat{{\hat n}}
\def\phat{{\hat p}}
\def\qhat{{\hat q}}
\def\rhat{{\hat r}}
\def\shat{{\hat s}}
\def\Ibar{{\bar I}}
\def\Jbar{{\bar J}}

\begin{document}
\begin{titlepage}
\setcounter{page}{0}
\begin{flushright}
  arXiv:1206.nnnn\\ NSF-KITP-12-101\\ UPR 1239-T
\end{flushright}
\vspace*{\stretch{1}}
\begin{center}
  \huge A class of Calabi-Yau threefolds as manifolds of $\SU(2)$ structure
\end{center}
\vspace*{\stretch{0.75}}
\begin{center}\hskip5.5pt  % need to fix this and center properly
  \large Michael B. Schulz\symbolfootnote[1]{mbschulz at brynmawr.edu}
\end{center}
\begin{center}
  \textit{Department of Physics, Bryn Mawr College\\
    Bryn Mawr, PA 19010, USA\\}
    \end{center}
\vspace*{\stretch{1}}
\begin{abstract}
  \normalsize A class of abelian fibered Calabi-Yau threefolds
  $\CX_{m,n}$ is shown yield $\SU(2)$ structure, in addition to the
  standard $\SU(3)$ holonomy.  Compactification of type~II string
  theory on a manifold in this class give a 4D effective supergravity
  theory in which the topology spontaneously breaks $\CN=4$ to $\CN=2$
  supersymmetry.  The breaking occurs at a scale hierarchically lower
  than the compactification scale when the $\IP^1$ base is large
  compared to the $T^4$ fiber.  We analyze the moduli space of $SU(2)$
  structure metrics of the $\CN=4$ theory and its restriction to the
  moduli space of Calabi-Yau metrics of the $\CN=2$ theory, showing
  that the latter agrees with the expectation computed from triple
  intersection numbers in the classical limit.  Finally, we analyze
  the twisted cohomology ring associated with the $SU(2)$ structure of
  $\CX_{m,n}$ and show that the breaking of $\CN=4$ to $\CN=2$ is
  conveniently summarized in the lifting cohomology classes as one
  passes to the standard cohomology ring, with massive modes
  persisting as torsion classes when the coupling is nonminimal.  The
  analysis is facilitated by the existence of explicit first-order
  metrics obtained by classical supergravity dualities.
   \end{abstract}
  \vspace*{\stretch{5}}
  \flushleft{18 June 2012}
%\flushleft{Draft: \today}
\end{titlepage}

%%%%%%%%%%%%%%%%%%%%%%%%%%%%%
%%%   Table of contents   %%%
%%%%%%%%%%%%%%%%%%%%%%%%%%%%%

\tableofcontents
\newpage

%%%%%%%%%%%%%%%%%%%%%%%%%%%
%%%   1. Introduction   %%%
%%%%%%%%%%%%%%%%%%%%%%%%%%%

\section{Introduction}
\label{sec:Intro}

Calabi-Yau manifolds are the bread and butter of string theory
compactifications.  From traditional heterotic models, to type IIB
flux compactifications, to IIA intersecting brane models based on
Calabi-Yau orientifolds, it is hard to imagine model building that is
not in some way related to Calabi-Yau
compactifications.\footnote{$G_2$ compactifications are no exception.
  They give rise to type IIA intersecting D6 brane models on
  Calabi-Yau orientifolds in the weak coupling limit, and to heterotic
  compactifications in the case that the $G_2$ manifold admits a $\K3$
  fibration.}  The study of Calabi-Yau mirror symmetry in the late 80s
and early 1990s ushered in a renaissance of cross fertilization
between mathematics and physics that continues to yield new insights,
and has brought us the inspiring edifice of topological string theory
in its closed and open forms. Given that Calabi-Yau manifolds have
been studied for more than two and a half
decades~\cite{Candelas:1985en}, it might seem unlikely that any novel
low hanging fruit would not already have been plucked.

The rise of flux compactifications a decade ago brought one such
development.  It was once thought that nonzero flux was forbidden on
energetic grounds, but we now know that that argument does not apply,
due to the existence of $\a'$ corrections and classically negative
tension objects such as orientifold planes and wrapped D-branes that
are well defined in the microscopic theory.  The flux can be thought
of as a discrete choice of data that \emph{gauges} the low energy
supergravity theory, coupling vectors and scalars that would otherwise
not interact.  It was in this context that the first (metastable) de
Sitter vacua~\cite{Kachru:2003aw} and the first compactifications with
all moduli stabilized~\cite{Denef:2005mm} were constructed.  In
another departure from purely geometric Calabi-Yau compactification,
discrete modifications of Calabi-Yau topology were considered, which
give rise to manifolds of $\SU(3)$ structure rather than $\SU(3)$
holonomy. (For one of many examples, see \cite{Tomasiello:2005bp}.)
Again, the discrete data gauges the low energy supergravity theory.
The general framework giving rise to 4D $\CN=2$ gauged supergravity in
type II was described in Refs.~\cite{Grana:2005ny,Grana:2006hr}.  Such
compactifications are in general nongeometric.  When a local geometric
description exists, the local compactification manifold $X$ seems to
be one of with generalized tangent bundle $(T+T^*)X$ of $\SU(3)\times
\SU(3)$ structure.

In this paper, we return from these exotic developments to the
familiar, purely geometric $\CN=2$ Calabi-Yau compactification of type
II string theory in the absence of flux, and ask whether there might
be a gauged supergravity of higher supersymmetry lurking the
Calabi-Yau compactification itself.  Indeed, a potential mechanism for
this was sketched in Ref.~\cite{Donagi:2008ht}.  It is possible that
in addition to the standard $\SU(3)$ holonomy of a Calabi-Yau
threefold, which gives rise to a covariantly constant spinor and low
energy $\CN=2$ supersymmetry of type II, there might be a richer
$\SU(2)$ structure, with twice the number of global spinors and low
energy gauged $\CN=4$ supergravity.  In this context, only one spinor
would be covariantly conserved with respect to the metric connection,
and the topology of the Calabi-Yau would break the $\CN=4$ to $\CN=2$
at sufficiently low energies.

We know of at least one context, where this \emph{must} be the case.
A simple model embodying many of the features of more realistic flux
compactifications is the type IIB $T^6/\IZ_2$ orientifold.  $\CN=2$
vacua of this model were studied in
Refs.~\cite{Kachru:2002he,Schulz:2002eh,Schulz:2004ub}, and it was
shown in Ref.~\cite{Schulz:2004tt} that these vacua are dual to
standard Calabi-Yau compactifications of type IIA string theory.  The
fluxes of these $\CN=2$ vacua are parametrized by two integers $m$ and
$n$ determining the NSNS and RR flux in type IIB, related to the
number $M$ of D-branes by $M = 16-4mn$.  The IIA Calabi-Yau duals
$\CX_{m,n}$ are Abelian surface fibrations ($T^4$ analogs of elliptic
fibrations), with $m,n$ determining the \emph{topology} of the
fibration over the $\IP^1$ base.  Since the fluxes spontaneously break
$\CN=4$ to $\CN=2$ in the type IIB dual, it must be the case that the
topology of the $\CX_{m,n}$ spontaneously breaks $\CN=4$ to $\CN=2$ in
the Calabi-Yau compactification.  Since $\CN=4$ requires twice the
number of spinors as $\CN=2$, the spinor bilinears yield a richer set
of tensors on $\CX_{m,n}$ than the K\"ahler form and holomorphic
3-form $J$ determined by its Calabi-Yau structure alone.  We have an
$\SU(2)$ structure on $\CX_{m,n}$ and a corresponding moduli space of
$\SU(2)$ structure metrics, of which the Calabi-Yau metrics form a
small subset.

The defining feature of $\SU(2)$ structure is restricted $\SU(2)$
structure group of the frame bundle of $\CX_{m,n}$, which need not
(and does not) coincide with the Riemannian holonomy group from the
metric connection.  Instead it corresponds to a connection with
torsion, and with this torsion comes a twisted differential operator,
from which one can compute a twisted cohomology ring.  Just as the
massless fields of a Calabi-Yau compactification are intimately
connected with the harmonic forms in one-to-one correspondence the
ordinary de Rham cohomology, the light fields of the $\CN=4$ theory
are connected in the same way with twisted harmonic forms and twisted
cohomology.  A subset of these light $\CN=4$ fields are the exact
massless fields of the $\CN=2$ theory, and likewise, a subset of the
twisted cohomology representatives are represent the small nontwisted
cohomology ring.

When nonminimal a flux is chosen in type IIB, some of the $\U(1)$ gauge
groups are coupled in such a way that all charged matter has $N$ times
the minimal unit of charge, and the resulting Higgs mechanism leaves a
$\IZ_N$ subgroup of the original $U(1)$ unbroken.  Correspondingly,
when nonminimal fibration data $(m,n)\ne(1,1)$ is chosen in defining
the topology of $\CX_{m,n}$, a discrete subset of the larger twisted
cohomology ring persists in the ordinary cohomology ring as torsion
classes.

In this paper, we study the $\SU(2)$ structure and Calabi-Yau geometry
of the family $\CX_{m,n}$ utilizing an approximate model for the
metric obtained by duality to the tree level supergravity description
of the type IIB $T^6/\IZ_2$ orientifold with $\CN=2$
flux~\cite{Schulz:2004tt}. (For similar first-order analyses in the
contact of K3 and F-theory, see
Refs.~\cite{Schulz:2012wu,Grimm:2012rg}.) This approximate description
is not the only one available.  Exact constructions were given in
Ref.~\cite{Donagi:2008ht}.  These results and their extensions will be
utilized in this paper as well.  However, the benefit of the
approximate description is that the metric, twisted and nontwisted
cohomology ring and harmonic forms, are substantially more accessible
than they would be using the tools of algebraic geometry, and
nevertheless suffice to yield the discrete topological data that one
would want in the exact description.

Prior work on $\SU(2)$ structure compactifications include a chain of
interesting investigations~\cite{Gauntlett:2003cy,Bovy:2005qq,
  ReidEdwards:2008rd,Triendl:2009ap,Louis:2009dq,Danckaert:2009hr,
  Spanjaard:2008zz, Danckaert:2011ju}, which were primarily concerned
with the low energy effective field theory as opposed to explicit
examples of manifolds of $\SU(2)$ structure.  In contrast, we will
focus on the geometry and differential topology of $\SU(2)$ structure,
in the explicit context of the family $\CX_{m,n}$, touching on
effective field theory only to the extent that it relates to moduli
spaces and the twisted versus ordinary cohomology.  The observation
that a Calabi-Yau manifold might furnish a manifold of $\SU(2)$
structure was made independently by Kashani-Poor, Minasian, and
Triendl, who presented preliminary results of their work at Strings
2011.  Their work~\cite{TriendlStringsTalk,KashaniPoorSU2} considers
the low energy effective field theory from compactification on the
Voisin-Borcea Calabi-Yau threefold $(\K3\times T^2)/\IZ_2$ fibered by
Enriques surfaces and familiar from Ref.~\cite{Ferrara:1995yx}.  These
authors also study certain generalities that the present work does
not, for example noting that the Euler characteristic must vanish as a
condition for the existence of the global 1-forms of $\SU(2)$
structure in 6D.

An outline of the paper is as follows.  

In Sec.~\ref{sec:Generalities} we introduce notation, and discuss
$\SU(2)$ and $\SU(3)$ structure generalities, for example, the
relation between global nowhere-vanishing spinors and invariant
tensors constructed from these spinors.  In the case of $\SU(3)$
structure we obtain the familiar $J$ and $\O$, while in the the
$\SU(2)$ structure the coframe bundle splits into 2D and 4D subbundles
$\CF^* = \CF_4^*+\CF_2^*$, with a pair of global nowhere-vanishing
1-forms trivializing a 2D subbundle of the frame bundle, together with
an almost hypercomplex structure of the 4D subbundle.  The torsion and
twisted exterior derivatives are introduced in this section.  We close
the section with a discussion of the case of interest: simultaneous
$\SU(3)$ holonomy and $\SU(2)$ structure.

Sec.~\ref{sec:ClassAbelianFibCY} gives a quick review of the
properties of the abelian surface fibered Calabi-Yau threefolds
$\CX_{m,n}$, and their origin via duality from the type IIB
$T^6/\IZ_2$ orientifold with $\CN=2$ flux~\cite{Schulz:2004tt}.  We
also sketch some of the features of the explicit constructions of
Ref.~\cite{Donagi:2008ht}.

Sec.~\ref{sec:FirstOrderSU2Met} focuses on $\CN=4$ description, in
which the $\CX_{m,n}$ viewed as manifolds of $\SU(2)$ structure.  We
discuss the space of approximate $\SU(2)$ structure metrics obtained
by duality from the $\CN=4$ theory, truncating to the tree level type
IIB supergravity description in the $T^6/\IZ_2$ dual.  We describe the
frame, harmonic forms, and moduli spaces of almost hypercomplex
structures on $\CF_4^*\CX_{m,n}$ and almost complex structures on
$\CF_2^*\CX_{m,n}$.  This section also defines the specific torsion
and twisted exterior derivatives for $\CX_{m,n}$.

Sec.~\ref{sec:FirstOrderCYMetric} turns to the lower energy theory in
which $\CN=4$ is spontaneously broken to $\CN=2$, and the remaining
massless degrees of freedom coincide with the standard description of
the $\CX_{m,n}$.  We give the restriction of the metric from the more
general $\SU(2)$ form described in the previous section, and write
down the K\"ahler form and holomorphic 3-form.  Restricting the
$\CN=4$ moduli space to the exact moduli of the $\CN=2$ Calabi-Yau
metric, we obtain an expression for the (self-mirror) moduli space
metric on the space of first-order Calabi-Yau metrics.  This metric
agrees with the classical (i.e., cubic prepotential) Calabi-Yau
metrics computed from the classical triple intersection form on
$\CX_{m,n}$, after a field redefinition.  The details of the field
redefinition and metric equivalence can be found in
App.~\ref{app:Kahler}.  We close this section with a discussion of
harmonic forms, and define a moduli independent basis of
$H_{2}(\CX_{m,n},\IR)$, which we identify with a basis from the exact
description of $\CX_{m,n}$ later in Sec.~\ref{sec:IntegerHomology}.

This brings us to Sec.~\ref{sec:TwistedVsOrdinary}, which relatively
short, and perhaps the most novel part of the paper.  In a basis
defined in Secs.~4 and 5, we compute the twisted and ordinary
cohomology rings of $\CX_{m,n}$.  The latter can be computed within
the former, and lifts a number of classes which become nonclosed or
exact with respect to the ordinary exterior derivatives.  When the
topological data is nonminimal, $(m,n)\ne(1,1)$, we find that remnants
of the lifted cohomology classes persist as torsion classes.  In the
final part of this section, we discuss the free part of the integer
homology from the point of view of explicit constructions, in order to
put earlier results on firmer ground and make precise statements about
how the moduli independent forms defined at the end of
Sec.~\ref{sec:FirstOrderCYMetric} relate to integer homology classes.
In App.~\ref{app:Intersections} we compute the double and triple
intersections of divisors, extending the results of
Ref.~\cite{Donagi:2008ht}.

Finally, in Sec.~\ref{sec:Conclusions}, we conclude, and discuss
relations to ongoing and future work.

%%%%%%%%%%%%%%%%%%%%%%%%%%%%%%%%%%%%%%%%%%%%%%%%%%%%%
%%%   2. SU(3) and SU(2) structure generalities   %%%
%%%%%%%%%%%%%%%%%%%%%%%%%%%%%%%%%%%%%%%%%%%%%%%%%%%%%

\section{$\SU(3)$ and $\SU(2)$ structure generalities}
\label{sec:Generalities}

Consider a six dimensional oriented Riemannian manifold $(\CX,g)$ with
vanishing second Stiefel-Whitney class $w_2\in H^2(\CX,\IZ_2)$, so
that $\CX$ admits a spin structure.\footnote{The spin structures are
  in one-to-one correspondence with elements of $H^1(\CX,\IZ_2)$, so
  there is a unique spin structure for simply connected $\CX$, but may
  be more than one otherwise.}  Let $x^m$ denote coordinates on $\CX$
and let $e^{\mhat}{}_n$ be a vielbein for $g_{mn}$,
\begin{equation}
  ds^2 = g_{mn} dy^m dy^n = \d_{\mhat\nhat} e^\mhat e^\nhat,
  \quad\text{where}\quad e^a = e^a{}_m dy^m.
\end{equation}
The Clifford algebra is
\begin{equation}
	\{\g_m,\g_n\} = 2g_{mn},
\end{equation}
where $\g_m = \g_\nhat e^\nhat{}_m$, and the $\g_\nhat$ are constant
gamma matrices satisfying $\{\g_\mhat,\g_\nhat\} = 2\d_{\mhat\nhat}$.
Since $\CX$ is oriented, we can define a volume form and chirality
operator by
\begin{align}
  \Vol_{(6)} &= e^{\hat1}\w e^{\hat2}\w e^{\hat2}\w e^{\hat4}\w
  e^{\hat5}\w e^{\hat6}  
  = \sqrt{g} dy^1\w dx^2\w dx^3\w dx^4\w dx^5\w dx^6,\\
  \g_{(6)} &= \frac1{6!}\e_{\mhat\nhat\phat\qhat\rhat\shat}
  \g^{\mhat\nhat\phat\qhat\rhat\shat}
  = \frac1{6!}(\Vol_{(6)})_{mnpqrs}\g^{mnpqrs},
\end{align}
where gamma matrices with multiple indices denotes antisymmetrized
products of gamma matrices: for example, $\g_{mn} =
\g_{[m}\g_{n]}=\half (\g_m\g_n-\g_n\g_m)$.

%%%%%%%%%%%%%%%%%%%%%%%%%%%%%%%%%%%%%%
%%%   2.1. SU(3) structure in 6D   %%%
%%%%%%%%%%%%%%%%%%%%%%%%%%%%%%%%%%%%%%

\subsection{$\SU(3)$ structure in 6D}
\label{sec:SU3}

%%%%%%%%%%%%%%%%%%%%%%%%%%%%%%%%%%%%%%%%%%%%%%%%%%%%%%
%%%   2.1.1. Spinors and SU(3) invariant tensors   %%%
%%%%%%%%%%%%%%%%%%%%%%%%%%%%%%%%%%%%%%%%%%%%%%%%%%%%%%

\subsubsection{Spinors and $\SU(3)$ invariant tensors}

Suppose further that there exists a global nowhere-vanishing spinor
$\chi$ of positive chirality, $\g\chi = \chi$, which we assume without
loss of generality to be normalized, $\chi^\dagger\chi = 1$.  Then,
$\CX$ has $\SU(3)$ structure.\footnote{A $d$-dimensional manifold $\CX$
  is said to have $G$-structure when the structure group of the frame
  bundle is reduced from $SO(d)$ to a subgroup $G$.}  That is, the
structure group of the frame bundle of $\CX$ is reduced from $SO(6)$
to $\SU(3)$, and $\chi$ determines $\SU(3)$ invariant tensors
\begin{align}
  J_m{}^n = i\chi^\dagger\g_m{}^n\chi, \label{eq:JSU3}\\
  \O_{mnp} = \chi^\dagger\g_{mnp}\chi^* \label{eq:OmegaSU3}. 
\end{align}
In writing the latter equation, we employ a Majorana convention, in
which the $\g_m$ are imaginary and Hermitian.  We assume this
convention throughout the paper.  From Fierz identities, it can be
shown that
\begin{gather}
  J_m{}^n J_n{}^p = -\d_m{}^p,\\
  \frac1{3!} J\w J\w J = \frac{i}8 \O\w\bar\O = \Vol_{(6)}.	
\end{gather}
Here, the first line implies that $J_m{}^n$ defines an almost complex
structure (ACS).  In the second line, the fundamental form $J=\half
J_{mn}dx^m\w dx^n$ is obtained by using the metric to lower an index
of the ACS, $J_{mn} = J_m{}^p\d_{pn}$.  Given a holomorphic
(antiholomorphic) frame $e^j$ ($e^\jbar$), we have
\begin{equation}
  J_j{}^k = i\d_j{}^k,\quad J_\jbar{}^\kbar = -i\d_\jbar{}^\kbar,
\end{equation}
from which the gamma matrices $\g^\jbar,\g_k$ annihilate $\chi$, and
the gamma matrices $\g^j,\g_\kbar$ act as raising operators.  Thus, we
have an isomorphism between holomorphic differential forms
$\o_{i_1\dots\i_p}$ and spinors $\o_{i_1\dots
  i_p}\g^{i_1\dots\i_p}\chi$.  The normalized negative chirality
spinor $\chi^*$ satisfies $\g_{ijk}\chi^* = \O_{ijk}\chi$.

%%%%%%%%%%%%%%%%%%%%%%%%%%%%%%%%%%%%%%%%%%%%%%%%%%%%%%%%%%%%%%%%%%%%%%%%%
%%%   2.1.2. Torsion, contorsion, and twisted exterioir derivatives   %%%
%%%%%%%%%%%%%%%%%%%%%%%%%%%%%%%%%%%%%%%%%%%%%%%%%%%%%%%%%%%%%%%%%%%%%%%%%

\subsubsection{Torsion, contorsion, and twisted exterior derivative}

When $\chi$ is covariantly constant with respect to the standard spin
connection $\nabla$, it is straightforward to show that $dJ$ and $d\O$
vanish.  Then, the manifold is Calabi-Yau and the standard connection
has $\SU(3)$ holonomy.  For more general $\SU(3)$ structure manifolds,
$\chi$~is conserved by a connection $\nabla^{(T)}$ with torsion,
\begin{equation}\label{eq:nablaT}
  \nabla^{(T)}_m\chi = (\nabla_m - \frac14\k_{mnp}\g^{np})\chi = 0,
\end{equation}
where $\k_{mnp}$ is the
contorsion~\cite{Gurrieri:2002wz}.\footnote{See
  Ref.~\cite{Nakahara:2003nw} Sec.~7.2.6 for a pedagogical discussion
  of torsion and contorsion, and Ref.~\cite{Joyce:2000} Sec.~2.6 for a
  discussion of intrinsic torsion.}  The contorsion tensor gives the
difference between the connection and the standard metric-connection,
while the torsion tensor $T_{mn}{}^p = \G_{mn}{}^p-\G_{nm}{}^p$ gives
the antisymmetric component of the connection.  The relation between
the two is $\k_{mnp} = \frac12(T_{mnp} - T_{npm} + T_{pmn})$.

As discussed in Refs.~\cite{Gurrieri:2002wz}, the contorsion can be
viewed as an $\so(6)$ Lie algebra valued 1-form, and decomposes as $\k
= \k^{\su(3)} + \k^0$, where $\k^0$ lies in $\so(6)/\su(3)$.  The
action of $\k^{\su(3)}$ on $\chi$ vanishes, since $\chi$ is an
$\SU(3)$ singlet.  So, it is actually only the intrinsic contorsion
$\k_0$ that contributes to Eq.~\eqref{eq:nablaT}.

When the torsion vanishes, there is no difference between the exterior
derivative and a ``covariant exterior derivative" obtained by
replacing partial derivatives with covariant derivatives.  However,
when the torsion is nonzero and one makes the same substitution, only
the symmetric part of the connection drops out, and the antisymmetric
torsion part remains.  The result is a twisted exterior derivative
$d^{(T)}$ defined by
\begin{equation}\label{eq:dtwisted}
  d^{(T)}\o = d\o -T^p\w\iota_p\o,
  \quad\text{where}\quad T^p = T_{mn}{}^p dx^m\w dx^n.
\end{equation}
Just as the deformation space of a Calabi-Yau metric is determined by
the $d$-cohomology ring $H^*(\CX,\IC)$, one expects the natural
deformation space of an $\SU(3)$ structure metric to be closely
related to the $d^{(T)}$-cohomology ring $H^*_{(T)}(\CX,\IC)$.  An
analogous expectation holds in context of $\SU(2)$ structures; we will
see how this expectation is realized in
Sec.~\ref{sec:TwistedVsOrdinary}.

%%%%%%%%%%%%%%%%%%%%%%%%%%%%%%%%%%%%%%
%%%   2.2. SU(2) structure in 6D   %%%
%%%%%%%%%%%%%%%%%%%%%%%%%%%%%%%%%%%%%%

\subsection{$\SU(2)$ structure in 6D}
\label{sec:SU2}

%%%%%%%%%%%%%%%%%%%%%%%%%%%%%%%%%%%%%%%%%%%%%%%%%%%%%%
%%%   2.2.1. Spinors and SU(2) invariant tensors   %%%
%%%%%%%%%%%%%%%%%%%%%%%%%%%%%%%%%%%%%%%%%%%%%%%%%%%%%%

\subsubsection{Spinors and $\SU(2)$ invariant tensors}

When there exist not one, but two global nowhere-vanishing
everywhere-independent positive chirality spinors on $\CX$, the
structure group of the frame bundle is further reduced to $\SU(2)$.
We assume without loss of generality that the two spinors are
normalized, $\chi_1^\dagger\chi_1 = \chi_2^\dagger\chi_2$, and
orthogonal, $\chi_1^\dagger\chi_2=0$.  The spinors determine a triple
of tensors $(J^\a)_a{}^b$, $\a=1,2,3$, and a pair of real 1-forms
$w^1,w^2$ on $\CX$,
\begin{equation}
  (J^\a)_a{}^b = \frac{i}2\chi^\dagger \s^\a\g_a{}^b\chi,\quad
  w^\a_a = \frac12\chi^T\s^\a\g_a\chi,
  \quad\text{where}\quad	
  \chi = \begin{pmatrix} \chi_1\\ \chi_2\end{pmatrix}.
\end{equation}
Here, $\s^\a$ are the $2\times2$ Pauli matrices, satisfying
$\s^\a\s^\b = \d^{\a\b} + i\e^{\a\b\g}\s^\g$.  The 1-form $w^3$
vanishes.  The condition that $w^1,w^2$ be real is a relative phase
convention on $\chi_1$ and $\chi_2$.  An~overall phase rotation of
$\chi$ acts as $SO(2)$ rotation on $w^1$ and $w^2$.  Lowering the
second index of $(J^\a)_a{}^b$, we obtain a triple of 2-forms $J^\a =
\half (J^\a)_{ab}e^a\w e^b$, which can be written
\begin{equation}\label{eq:SU2InvTensors}
  J^1 = j^1,\quad J^2 = j^2,\quad J^3 = j^3 + w^1\w w^2,
\end{equation}
with the following interpretation.

%%%%%%%%%%%%%%%%%%%%%%%%%%%%%%%%%%%%%%%%%%%%%%%%%%%%%%%%%%%%%%%%%%%%%%%%%
%%%   2.2.2. Bundle decomposition and almost hypercomplex structure   %%%
%%%%%%%%%%%%%%%%%%%%%%%%%%%%%%%%%%%%%%%%%%%%%%%%%%%%%%%%%%%%%%%%%%%%%%%%%

\subsubsection{Bundle decompositions and almost hypercomplex
  structure}

The coframe bundle $F^*\CX$ splits as the sum of a 4D subbundle
$F_4^*\CX$ of structure group $\SU(2)$ and 2D trivial subbundle
$F_2^*\CX$.  The $(j^\a)_\a{}^\b$ give a triple of almost complex
structures satisfying the quaternionic algebra, on $F_4^*\CX$.  The
forms $w^1$ and $w^2$ trivialize $F_2^*\CX$.  On $F_2^*\CX$, there is
a single almost complex structure, with holomorphic 1-form $v_1+iv_2$.

Since the frame bundle splits, the spinor bundle correspondingly
factorizes as a $\Spin(4)$ subbundle times $\Spin(2)$ subbundle.  The
$\Spin(4)$ subbundle, of structure group $\SU(2)$, admits global
nowhere-vanishing spinors $\chi^{(4)}_\pm$, where the subscripts give
the $\g_{(4)} = \g_{1234}$ chirality.  The $\Spin(2)$ bundle, of
trivial structure group, admits global nowhere-vanishing spinors
$\chi^{(2)}_\pm$, where the subscripts give the $\g_{(2)} = i\g_{56}$
chirality.  Here, components of gamma matrices are with respect to a
frame basis that respects the split $F\CX = F_4\CX+F_2\CX$.  The pair
of global nowhere-vanishing positive chirality spinors is $\chi_1 =
\xi_+\otimes\z_+$ and $\chi_2 = \xi_-\otimes\z_-$.

Introducing 4D and 2D gamma matrices, $\tilde\g_\a$ and $\tilde\g_\r$,
respectively,
\begin{equation}
  \{\tilde\g_\a,\tilde\g_\b\} = 2\d_{\a\b},\quad \a,\b=1,2,3,4,
  \quad\text{and}\quad 
  \{\tilde\g_\r,\tilde\g_\s\} = 2\d_{\r\s},\quad \r,\s=5,6,
\end{equation}
we can write the 6D gamma matrices as $\g_\r = 1\otimes\tilde\g_\r$
and $\g_\a =\g_{(2)}(\tilde\g_\a\otimes1)$.  Here, $\g_{(4)} =
\tilde\g_{(4)}\otimes1$, $\g_{(2)}= 1\otimes\tilde\g_{(2)}$, and
$\g_{(6)} = \tilde\g_{(4)}\otimes\tilde\g_{(2)}$, where
$\tilde\g_{(4)} \tilde\g_{1234}$ and $\tilde\g_{(2)} =
i\tilde\g_{56}$.  The previous definitions are then equivalent to
\begin{equation}
   (j^\a)_\a{}^\b = \frac{i}2\xi^\dagger \s^\a\tilde\g_\a{}^\b\xi
   \quad\text{and}\quad
  (w^1+iw^2)_\r = \z_+^T\tilde\g_\r\z_-,
  \quad\text{where}\quad	
  \xi = \begin{pmatrix} \xi_+\\ \xi_-\end{pmatrix}.
\end{equation}
This makes it clear that the 1-forms $w^1$ and $w^1$ are sections of
$F_2^*\CX$ and the the 2-forms $j^\a$ (with lowered indices) are
sections of $\w^2 F_4^*\CX$.

In the Majorana convention assumed in Sec.~\ref{sec:SU3}, $\g_a$ and
$\g_{(6)}$ and $\g_{(2)}$ are imaginary and Hermitian, while
$\g_{(4)}$ is real and Hermitian.  Thus, complex conjugation of a 6D
spinor reverses its $\g_{(6)}$ and $\g_{(2)}$ chiralities, leaving its
$\g_{(4)}$ chirality unchanged.  Similarly, $\tilde\g_\a$ is real and
Hermitian, while $\tilde\g_\r$ is imaginary and Hermitian.  The
simplest compatible phase convention for $\xi_\pm$ and $\z_\pm$ is
$\xi_\pm^* = \xi_\pm$ and $\z_\pm^* = \z_\mp$.  This leads to a
variety of alternative expressions for $j^\a$ and $w^1+iw^2$.  For
example, of relevance below, we note that
\begin{equation}\label{eq:JSU3fromSU2}
  (j^3)_\a{}^\b = i\xi_+^\dagger\tilde\g_\a{}^\b\xi_+,
  \quad\text{and}\quad
  (J^3)_\a{}^\b = i\chi_1^\dagger\g_a{}^\b\chi_1.
\end{equation}

From Fierz identities, the almost complex structures $(j^\a)_\a{}^\b$
on $F_4^*\CX$ satisfy
\begin{equation}
  (j^\a)_\a{}^\g(j^\b)_\g{}^\b = -\d^{AB} - \e^{ABC}(j^C)_\a{}^\b.
\end{equation}
This is the quaternionic algebra up to an expected minus sign on the
second term.\footnote{the transpose of $(j^A)_\a{}^\b$ acts on the
  frame rather than coframe bundle (i.e., tangent rather than
  cotangent bundle) and satisfies the quaternionic algebra with the
  standard sign conventions.}  We also have
\begin{equation}
  J^A\w J^B = \d^{AB}\Vol_4,\quad
  w^1\w w^2 = \Vol_2,
  \quad\text{and}\quad
  \Vol_6 = \Vol_4\w\Vol_2,
\end{equation}
where $\Vol_4$ and $\Vol_2$ are the volume forms associated to
$F_2^*\CX$ and $F_4^*\CX$, respectively.

%%%%%%%%%%%%%%%%%%%%%%%%%%%%%%%%%%%%%%%%%%%%%%%%%%%%%%%%%%%%%%%%%%%%%%%%
%%%   2.2.3. Torsion, contorsion, and twisted exterior derivatives   %%%
%%%%%%%%%%%%%%%%%%%%%%%%%%%%%%%%%%%%%%%%%%%%%%%%%%%%%%%%%%%%%%%%%%%%%%%%

\subsubsection{Torsion, contorsion, and twisted exterior derivative}

When $\chi_1$ and $\chi_2$ are covariantly constant with respect to
the standard spin connection $\nabla$, it is straightforward to show
that $dJ$ and $d(w^1+iw^2)$ vanish.  Then, the manifold $\CX$ has
$\SU(2)$ holonomy, and globally factorizes as $\K3\times T^2$.  More
generally, as in the previous section, the spinors are instead
covariantly constant with respect to a connection $\nabla^{(T)}$ with
torsion,
\begin{equation}
  \nabla^{(T)}_m\chi_{1,2} = (\nabla_m + \frac14\k_{mnp}\g^{np})\chi_{1,2} = 0.
\end{equation}
As in the discussion of the previous section, we then expect a
relation between the twisted cohomology defined by $d^{(T)}$
(cf.~Eq.~\eqref{eq:dtwisted}) and the deformation space of $\SU(2)$
structure metrics.

%%%%%%%%%%%%%%%%%%%%%%%%%%%%%%%%%%%%%%%%%%%%%%%%%%%%%%%
%%%   2.2.4. SU(3) structure from SU(2) structure   %%%
%%%%%%%%%%%%%%%%%%%%%%%%%%%%%%%%%%%%%%%%%%%%%%%%%%%%%%%

\subsubsection{$\SU(3)$ structure from $\SU(2)$ structure}

Finally, since $\SU(2)\subset \SU(3)$, we should be able to reproduce
the results of Sec.~\ref{sec:SU3} as a special case of this section,
simply forgetting about $\chi_2$.  This is indeed the case, and we
find
\begin{equation}\label{eq:JOfromSU2}
  J = j^3 + w^1\w w^2,\quad
  \O = (j^1+ij^2)\w(w^1+iw^2).
\end{equation} 
The first relation follows from Eqs.~\eqref{eq:JSU3} and \eqref{eq:JSU3fromSU2}, and the second follows from $\g_{\a\b\r} = \tilde\g_{ab}\otimes\tilde\g_\r$.

%%%%%%%%%%%%%%%%%%%%%%%%%%%%%%%%%%%%%%%%%%%%%%%%%%%%%%%%%%%%%%%%
%%%   2.3. Manifolds of SU(2) structure and SU(3) holonomy   %%%
%%%%%%%%%%%%%%%%%%%%%%%%%%%%%%%%%%%%%%%%%%%%%%%%%%%%%%%%%%%%%%%%

\subsection{Manifolds of $\SU(2)$ structure and $\SU(3)$ holonomy}
\label{sec:SU3SU2}

There is an interesting intermediate case, in which the contorsion
annihilates one of the two spinors of Sec.~\ref{sec:SU2}, say
$\chi_1$, but not the other, so that
\begin{equation}
  \nabla^{(T)}\chi_1 = \nabla^{(T)}\chi_2 = 0,
  \quad\text{and}\quad 
  \nabla\chi_1 = 0, \quad \nabla\chi_2\ne0.
\end{equation}
In this case, the same manifold $\CX$ is a Calabi-Yau manifold of
$\SU(3)$ holonomy (with spinor~$\chi_1)$, and also a manifold of
$\SU(2)$ structure (with spinors $\chi_1,\chi_2$).  The moduli space
of Calabi-Yau metrics is a proper subspace of the moduli space of
$\SU(2)$ structure metrics, and the ordinary de~Rham cohomology ring
$H^*(\CX,\IR)\cong\Harm(\CX)$ is a proper subring of the twisted
de~Rham cohomology ring $H^*_{(T)}(\CX,\IR)\cong\Harm_{(T)}(\CX)$.

In this paper, we analyze a class of Calabi-Yau manifolds for which a
stronger statement holds, in \emph{integer} cohomology.  The
Calabi-Yau manifolds are abelian fibrations over base $\CB=\IP^1$, and
the torsion can be interpreted as a map $T\colon
H_{(T)}^p(\CB,R^q\pi_*\IZ)\to
H_{(T)}^{p+1}(\CB,R^q\pi_*\IZ)$.\footnote{Here,
  $H_{(T)}^p(\CB,R^q\pi_*\IZ)$ is roughly the cohomology subgroup of
  degree $p$ on the base and degree $q$ on fiber.}  We obtain the
cohomology ring $H^*(\CX,\IZ)$ by computing the $d$-cohomology on a
space of representatives of $H_{(T)}^*(\CX,\IZ)$.  Thus, in a
spectral-sequence-like sense, $H_{(T)}^*(\CX,\IZ)$ gives a penultimate
approximation to $H^*(\CX,\IZ)$.  This is similar to the logic of
Ref.~\cite{Tomasiello:2005bp}.  There, the twisted homology ring of an
$\SU(3)$ structure manifold appeared as an approximation to its
ordinary cohomology ring, within a spectral sequence.\footnote{It so
  happens that the twisted cohomology ring in the example of
  Ref.~\cite{Tomasiello:2005bp} coincides with the ordinary cohomology
  ring of the mirror of the quintic Calabi-Yau 3-fold.}  Here, the
twisted cohomology of an $\SU(2)$ structure manifold appears as an
approximation to its ordinary cohomology ring.

%%%%%%%%%%%%%%%%%%%%%%%%%%%%%%%%%%%%%%%%%%%%%%%%%%%%%%%%%%%%%%%%%%%%%%
%%%   3. A class of abelian fibered Calabi-Yau manifolds X_{m,n}   %%%
%%%%%%%%%%%%%%%%%%%%%%%%%%%%%%%%%%%%%%%%%%%%%%%%%%%%%%%%%%%%%%%%%%%%%%

\section{A class of abelian fibered Calabi-Yau manifolds $\CX_{m,n}$}
\label{sec:ClassAbelianFibCY}

We focus on the class of Calabi-Yau 3-folds $\CX_{m,n}$ that were
studied in Ref.~\cite{Schulz:2004ub} (via duality) and
Ref.~\cite{Donagi:2008ht} (intrinsically).  The properties that will
be relevant for us here are:

\begin{enumerate}
\item $\CX_{m,n}$ is an abelian surface $(T^4)$ fibration over
  $\IP^1$, with $8+M$ singular fibers, $M\ge0$.
\item The Hodge numbers of $\CX_{m,n}$ are $h^{11}=h^{21} =M+2$, where
  the $M$ and the positive integers $m,n$, are constrained by
  $M+4mn=16$.
\item The polarization of the abelian fiber is $(\mbar,\nbar) =
  (m,n)/\gcd(m,n)$.  This means that the K\"ahler form on the abelian
  fiber is proportional to $\mbar dy^1\w dy^2 + \nbar dx^3 + dx^4$, a
  positive integer form that can be used to define a projective
  embedding.
\item The generic Mordell-Weil lattice of sections (mod torsion) of
  $\CX_{m,n}$ is $D_M$.  Here, generic means that all singular fibers
  are topologically I$_1$ times an elliptic curve ($T^2$).
\item The generic Mordell-Weil subgroup of torsion sections is
  $\MW_\tor=\IZ_m\times\IZ_m$.
\item The fundamental group is $\pi_1=\IZ_n\times\IZ_n$.
\item In a convenient basis, the nonzero intersection numbers are
\begin{equation}\label{eq:AHEintersections}
  H^2\cdot A = 2\mbar\nbar,\quad H\cdot\CE_I\cdot\CE_J = -\mbar\d_{IJ},
\end{equation}
where $A$ is the class of the abelian fiber.  This basis is not quite
integral.  Rather, $H$ and $\CE_I$ are moduli dependent.  However, for
the appropriate choice of moduli, the forms
$H-\frac{\mbar^2\nbar}{6}A$ and $\mbar\nbar\CE_I$ give half-integer
classes in $H(\CX_{m,n},\half \IZ)$, while $A$,
$H-\frac{\mbar^2\nbar}{6}A\pm\CE_I$ and $\mbar\nbar(\CE_I-\CE_J)$ give
integer classes in $H(\CX_{m,n},\IZ)$.  This is discussed in
Sec.~\ref{sec:RelationToAHE}.
\item The second Chern class of $\CX_{m,n}$ is the sum of the singular
  loci of singular fibers ($8+M$ elliptic curves).  Its only nonzero
  intersection with $H,A,\CE_I$ is
\begin{equation}
  H\cdot c_2 = 8+M.
  \end{equation}
\end{enumerate}
A few additional observations will be useful.  Since $\pi_1$ is
abelian, $H_1(\CX_{m,n},\IZ)=\pi_1(\CX_{m,n})= \IZ_n\times\IZ_n$,
which can be shown to be generated by the classes of two $S^1$'s in
the generic fiber.  In addition to the group $\MW_\tor =
\IZ_m\times\IZ_m$ of torsion sections, the $T^2$ product of the two
torsion $S^1$'s is a $\IZ_{\gcd(m,n)}$ torsion class, where
$\gcd(m,n)$ is the greatest common divisor of $m$ and $n$.  (It is
completely vertical, so it is not a section in $\MW_\tor$.)  This
exhausts all torsion 2-cycles.  Therefore, the complete integer
homology ring is
\begin{equation}\label{eq:CompleteIntHomRing}
  \begin{split}
  H_0(\CX_{m,n},\IZ) &= \IZ,\\
  H_1(\CX_{m,n},\IZ) &= \IZ_n\times\IZ_n,\\
  H_2(\CX_{m,n},\IZ) &= \IZ^{M+2}\times\IZ_m\times\IZ_m\times\IZ_{\gcd(m,n)},\\
  H_3(\CX_{m,n},\IZ) &= \IZ^{2(M+3)}\times\IZ_m\times\IZ_m\times\IZ_{\gcd(m,n)},\\
  H_4(\CX_{m,n},\IZ) &= \IZ^{M+2}\times\IZ_n\times\IZ_n,\\
  H_5(\CX_{m,n},\IZ) &= \emptyset,\\
  H_6(\CX_{m,n},\IZ) &= \IZ.
  \end{split}
\end{equation}

To arrive at this list, the free part follows from the Hodge numbers,
and the torsion part is obtained from $H_1^\tor = \IZ_n\times\IZ_n$
and $H_2^\tor = \IZ_m\times\IZ_m\times\IZ_{\gcd(m,n)}$ as follows.
Poincar\'e duality implies that $H^\tor_p = H_\tor^{6-p}$, and the
Universal Coefficient Theorem implies that $H^\tor_p = H_\tor^{p+1}$.
Thus, on general grounds, for a connected 6D orientable manifold,
\begin{equation}
  \begin{split}
    & H_1^\tor = H_4^\tor\quad (= H^2_\tor = H^5_\tor),\\
    & H_2^\tor = H_3^\tor\quad (= H^3_\tor = H^4_\tor),
  \end{split}
\end{equation}
along with $H_0^\tor = H_5^\tor = H_6^\tor = 0$ ($= H^1_\tor = H^6_\tor = H^0_\tor$).
For $\CX_{m,n}$, these torsion groups become
\begin{equation}\label{eq:TorsionHom}
  \begin{split}
    & H_1^\tor = H_4^\tor\quad (= H^2_\tor = H^5_\tor)\quad = \quad\IZ_n\times\IZ_n,\\
    & H_2^\tor = H_3^\tor\quad (= H^3_\tor = H^4_\tor)\quad = \quad\IZ_m\times\IZ_m\times\IZ_{\gcd(m,n)},
  \end{split}
\end{equation}

%%%%%%%%%%%%%%%%%%%%%%%%%%%%%%%%%%%%%%%%%%%%%%%%%%%%%
%%%   3.1. Intrinsic definitions and properties   %%%
%%%%%%%%%%%%%%%%%%%%%%%%%%%%%%%%%%%%%%%%%%%%%%%%%%%%%

\subsection{Intrinsic definitions and properties}

Two constructions of $\CX_{m,n}$ were given in
Ref.~\cite{Donagi:2008ht}.  The first construction presented the
explicit monodromy matrices of the $M+8$ singular fibers, and
described the homology via the algebra of $(p,q,r,s)$ string junctions
on the $\IP^1$ base.  The second construction, for the special case
$m=n=1$, realized $\CX_{1,1}$ as the relative Jacobian of a genus-2
fibration over $\IP^1$.  We now summarize this construction, as it
will provide us with a useful intuition for the general case.

First, recall that an abelian variety is projective variety that is
also an abelian group.  It is a complex torus $T^{2d}\cong
\IC^d/\Lambda$, for some lattice $\L$, with group addition and
additive identity inherited from $\IC^2$.  The fact that it is a
projective variety implies that the K\"ahler form is proportional to a
Hodge form~\cite{GandH}
\begin{equation}
  dx^1\w dx^2 + \d_1 dx^3\w dx^4 + \dots \d_g dx^{2g-1}\w dx^{2g},
\end{equation}
with $\d_i$ divisible by $\d_{i-1}$.  The Poincar\"e dual homology
class is represented by a Hodge divisor.  The case $g=1$ gives an
elliptic curve and the case $g=2$ gives an abelian surface.  The
$g$-tuple $(1,\d_1,\dots,\d_g)$ is known as the polarization of the
abelian variety.

The Jacobian of a genus-$g$ curve $C_g$ is the space of degree zero
line bundles on $C_g$, modulo linear equivalence.  It is an abelian
variety, which can be seen as follows.  $\Jac(C_g)$ can be thought of
as the space of Wilson lines, i.e., a $U(1)$ phase for each of the
independent 1-cycles generating $H_1(C_g,\IZ)=\IZ^{2g}$.  Thus
$\Jac(C_g)\cong T^{2g}$.  For a genus-2 curves, the Jacobian is an
abelian surface ($T^4$).  In general, the space of degree $d$ line
bundles over a space $\CX$ is referred to as the Picard group
$\Pic^d(\CX)$.  Thus $\Jac(C_g) = \Pic^0(C_g)$.  Jacobian varieties
are always principally polarized, meaning $\d_i=1$ for $i=1,\dots g$.

Given a surface $S$ that is fibered by genus-$g$ curves over a base
curve $\CB$, the relative Jacobian $\Pic^0(S/\CB)$, is obtained from
$S$ by replacing each fiber by its Jacobian.  This gives an abelian
fibration over $\CB$.  For $g=2$ and $\CB=\IP^1$, we obtain an abelian
surface fibration over $\IP^1$.  For the appropriate choice of $S$, it
was shown in Ref.~\cite{Donagi:2008ht}, that $\CX_{1,1} =
\Pic^0(S/\CB)$.

For $\CX_{1,1}$, we have $M=12$, and the homology group
$H_4(\CX_{1,1},\IZ)$ is generated by the abelian fiber $A$ together
with the $2M$ theta divisors $\Theta_I,\Theta'_I$, for $I=1,\dots,M$.
(See the discussion in App.~\ref{app:Intersections}.)  It can be shown
that $[\Theta_I+\Theta'_I]=[D]$, independent of $I$, so this indeed
gives $\dim(H^4) = M+2$.  The relation to the basis with convenient
intersections~\eqref{eq:AHEintersections} is $[\CE_I] =
\half[\Theta_I-\Theta'_I]$ and $[H] = \half[D] - \frac16[A]$, where
$[D] = [\CE_J +\CE'_J]$.

The generalizations of these statements for the Calabi-Yau manifold
$\CX_{m,n}$ can be found in Sec.~\ref{sec:IntegerHomology} and
App.~\ref{app:Intersections}.

%%%%%%%%%%%%%%%%%%%%%%%%%%%%%%%%%%%%%%%%%%%%%%%%%%%%%%%%%%%
%%%   3.2. Duality to type IIB orientifolds with flux   %%%
%%%%%%%%%%%%%%%%%%%%%%%%%%%%%%%%%%%%%%%%%%%%%%%%%%%%%%%%%%%

\subsection{Duality to type IIB orientifolds with flux}
\label{sec:DualityToIIB}

%%%%%%%%%%%%%%%%%%%%%%%%%%%%%%%%%%%%%%%%%%%%%%%%%%%%%
%%%   3.2.1. The duality map leading to X_{m,n}   %%%
%%%%%%%%%%%%%%%%%%%%%%%%%%%%%%%%%%%%%%%%%%%%%%%%%%%%%

\subsubsection{The duality map leading to $\CX_{m,n}$}

The family of Calabi-Yau manifolds $\CX_{m,n}$ was first studied in
Ref.~\cite{Schulz:2004ub} in the context of a duality map relating the
simplest type IIB flux compactifications---toroidal orientifold with
flux---to purely geometric Calabi-Yau compactifications of type IIA
string theory.  As discussed in Ref.~\cite{Schulz:2004ub}, the type
IIB $T^6/\IZ_2$ orientifold with the choice of $\CN=2$
flux~\cite{Kachru:2002he}
\begin{equation}
  \begin{split}
    F_{(3)}/\bigl((2\pi)^2\a'\bigr)
    &= 2m\bigl(dx^4\w dx^6 + dx^5\w dx^7)\w dx^9,\\
    H_{(3)}/\bigl((2\pi)^2\a'\bigr)
    &= 2n\bigl(dx^4\w dx^6 + dx^5\w dx^7)\w dx^8,
  \end{split}
\end{equation}
is dual to type IIA compactified on $\CX_{m,n}$.  The duality map has
two steps: (i) T-duality of $T^3(x^4x^5x^9)$, which give a type IIA
orientifold with $M$ D6-branes and eight O6 planes, followed by (ii)
an M-theory ``9,10-flip.'' The 9,10-flip lifts the type IIA background
to M-theory to give a new $x^{10}$ circle, and then compactifies on
$x^9$ to return to type IIA string theory in 10 dimensions.  The
M-theory geometry resulting from the lift of the IIA orientifold is
$\IR^{1,3}(x^0x^1x^2x^3)\times\CX_{m,n}(x^4x^5x^6x^7x^8x^{10})\times
S^1(x^9)$, where $\CX_{m,n}$ is an abelian surfaces fibration over
$\IP^1(x^6x^7)$, i.e., a holomorphic $T^4$ fibration with additional
structure (a zero section, addition of sections, and a theta divisor).
The $S^1(x^{10})$ is contained in $T^4(x^4x^5x^8x^{10})$.  From here,
we compactify on $S^1(x^9)$ to obtain type IIA string theory on
$\IR^{1,3}(x^0x^1x^2x^3)\times \CX_{m,n}(x^4x^5x^6x^7x^8x^{10})$.

%%%%%%%%%%%%%%%%%%%%%%%%%%%%%%%%%%%%%%
%%%   3.2.2. Analogy to K3 x T^2   %%%
%%%%%%%%%%%%%%%%%%%%%%%%%%%%%%%%%%%%%%

\subsubsection{Analogy to $\text{\rm K3}\times T^2$}

The 3-fold $\CX_{m,n}$ can be thought of as a twisted analog of
$\K3\times T^2$, with fewer than 16 exceptional divisors.  Recall that
an elliptic K3 is an elliptic fibration over $\IP^1$, generically with
24 I$_1$ Kodaira fibers.  Thus $\K3\times T^2$ is trivially a $T^4$
fibration over $\IP^1$, with a $T^2\subset T^4$ factorizing.  In
$\CX_{m,n}$, the $T^4$ no longer factorizes, and the number of
singular fibers is reduced from 24 to $M+8$, where $M = 16-4mn$.  In
the $T^6/\IZ_2$ dual, $M$ is the number of D3-branes.  The Hodge
numbers of $\CX_{m,n}$ are $h^{11} = h^{21} = M+2$, with possible
values $M+2=2,6,10,14$.

%%%%%%%%%%%%%%%%%%%%%%%%%%%%%%%%%%%%%%%%%%%%%%%%%%%%%%%%%%%%%%
%%%   3.2.3. Supersymmetry, spinors, and SU(2) structure   %%%
%%%%%%%%%%%%%%%%%%%%%%%%%%%%%%%%%%%%%%%%%%%%%%%%%%%%%%%%%%%%%%

\subsubsection{Supersymmetry, spinors, and $\SU(2)$ structure}

Just as choice of flux in the dual type IIB $T^6/\IZ_2$ orientifold
spontaneously breaks $\CN=4$ to $\CN=2$ supersymmetry, the
\emph{topology} of $\CX_{mn}$ spontaneously breaks $\CN=4$ to $\CN=2$
supersymmetry in the dual type IIA compactification on $\CX_{m,n}$.
When the $\IP^1$ base of $\CX_{m,n}$ is large compared to the $T^4$
fiber, there is a hierarchy of scales, and it is meaningful to
describe this supersymmetry breaking in the low energy effective field
theory.  In the absence of flux in IIB, the $\CN=4$ supersymmetry is
unbroken, and the dual IIA compactification manifold is $\K3\times
T^2$.

The doubled ($\CN=4$) supersymmetry of $K3\times T^2$ relative to the
$\CN=2$ of a Calabi-Yau 3-fold follows from the fact that the latter
admits one covariantly constant positive chirality spinor, whereas the
former admits two: $u^\K3_+\otimes u^{T^2}_+$ and $u^\K3_-\otimes
u^{T^2}_-$.  Here, $u^X_\pm$ is a $\pm$ chirality spinor on the space
$X$.  Orientation reversal on $X$ reverses the chirality of $X$.  For
$X=T^2$, this can be implemented by complex conjugation, and for $X=$
an elliptic K3, it can be implemented by complex conjugation of the
elliptic fiber.  Thus, viewing $\K3\times T^2$ as a $T^4$ fibration
over $\IP^1$, the two positive chirality spinors are related by
complex conjugation of the $T^4$ fiber.

The last paragraph carries over to $\CX_{m,n}$, with one modification.
Complex conjugation of the $T^4$ fiber of $\CX_{m,n}$ indeed relates
the standard covariantly constant positive chirality spinor to another
global positive chirality spinor.  However, complex conjugation of the
fiber alone ``breaks'' the complex structure.  The complex conjugate
$T^4$ fibration is no longer a holomorphic $T^4$ fibration over
$\IP^1$, due to the $m,n$ dependent monodromies about singular fibers.
Correspondingly, the new positive chirality spinor is not covariantly
constant with respect to the standard spin connection.  It together
with its negative chirality counterpart generates a spontaneously
broken $\CN=2$ supersymmetry in $\CN=4$.  On the other hand, this new
spinor \emph{is} covariantly constant relative to a connection with
torsion.  The pair of positive chirality spinors (together with their
negative chirality counterparts) gives rise to an $\SU(2)$ structure
on $\CX_{m,n}$.

%%%%%%%%%%%%%%%%%%%%%%%%%%%%%%%%%%%%%%%%%%%%%%%%%%%%%%%%%%%%%%%%%
%%%  4. First-order SU(2) structure metric and moduli space   %%%
%%%%%%%%%%%%%%%%%%%%%%%%%%%%%%%%%%%%%%%%%%%%%%%%%%%%%%%%%%%%%%%%%

\section{First-order $\SU(2)$ structure metric and moduli space}
\label{sec:FirstOrderSU2Met}

The type IIB $T^6/\IZ_2$ orientifold with
flux~\cite{Kachru:2002he,Schulz:2002eh} gives an $\CN=4$ low energy
supergravity theory valid below the compactification scale $1/R$.  We
assume this scale to be hierarchically smaller than string
scale.\footnote{The tree level supergravity theory is no-scale model,
  in which the scale $1/R$ is not fixed.}  The fluxes parametrize a
gauging of this $\CN=4$ supergravity theory known as a flat
gauging~\cite{Andrianopoli:2002mf}, which spontaneously breaks the
$\CN=4$ to $\CN=2$ at an energy scale $\a'/R^3$.  Intuitively, this
energy scale arises since the periods of the fluxes over 3-cycles of
volume $\sim R^3$ are quantized in units $\sim\a'$.

At energies $\a'/R^3 < E < 1/R$, the light scalars in the closed
string sector are the 21 metric moduli of $T^6$, 15 moduli from the RR
4-form axions $C_{(4)}$ with purely internal indices, and the
axion-dilaton $C_{(0)}+ ie^{-\phi}$.  In addition, there are $6M$
scalars from the $T^6$ positions of $M$ D3-branes (plus images).  At
energies below $\a'/R^3$, 10 metric moduli and 6 of the $C_{(4)}$
axions remain, together with all $6M$ D3 moduli.

When this is mapped to the type IIA compactification on $\CX_{m,n}$,
the same $\CN=4$ low energy supergravity theory is valid below the
compactification scale $1/V_X{}^{1/6}$.  The \mbox{$\CN=4$} theory is
spontaneously broken to $\CN=2$ at the scale $1/V_{\IP^1}{}^{1/2}$ of
the $\IP^1$ base.  The light scalars of the $\CN=4$ theory are $13+3M$
metric moduli $7+M$ $B$-field moduli, and the dilaton.  In the $\CN=2$
theory, the remaining light scalars are $6+3M$ metric moduli, $2+M$
\mbox{$B$-field} moduli, and the dilaton.

%%%%%%%%%%%%%%%%%%%%%%%
%%%   4.1. Metric   %%%
%%%%%%%%%%%%%%%%%%%%%%%

\subsection{Metric}
\label{sec:SU2Metric}

Truncating to the tree level supergravity description of $T^6/\IZ_2$
orientifold in type IIB, the chain of classical supergravity dualities
discussed in Sec.~\ref{sec:DualityToIIB} gives an approximate
description of $\CX_{m,n}$ based on a ``first-order'' $\CN=4$ metric
of the form
\begin{equation}\label{eq:Neq4Metric}
  ds^2_{\CN=4} =   
  \sqrt{2V_4}\bigl(\D^{-1}Z G_{\a\b} dx^\a dx^\b + \D Z^{-1}(dx^4+A)^2\bigr)
%  +\frac{V_2}{\im\t_2}|\eta^1+\t_2\eta^2|^2.
  + g_{\r\s}\eta^\r\eta^\s.
\end{equation}
Here, the quantities $\eta^\r$, for $\r=1,2$, are global 1-forms on
$\CX_{m,n}$ defined by
\begin{equation}
  \begin{split}
    \eta^1 = dy^1 + A^1,
    &\quad\text{where}\quad F^1=dA^1 = 2n dx^1\w dx^3,\\
    \eta^2 = dy^2 + A^2,
    &\quad\text{where}\quad F^2=dA^2 = 2n dx^2\w dx^3.\\
  \end{split}
\end{equation}
The coordinates $x^1,x^2,x^3,x^4,y^1,y^2$ have periodicity 1 and $x$
coordinates are further identified under the involution
\begin{equation}\label{eq:Involution}
  \CI_4\colon\ (x^\a,x^4)\mapsto(-x^\a,-x^4).
\end{equation}

This metric can be thought of as the twisted product of a 4D
Gibbons-Hawking metric in the $x^1,\dots,x^4$ directions (quotiented
by $\IZ_2$) and a 2D torus metric in the $y^1,y^2$ directions.  As we
will see, the corresponding 4D $+$ 2D split of the frame bundle of
$\CX_{m,n}$ realizes the $\SU(2)$ structure.

For $m=n=0$, this metric reduces to that of $\K3\times T^2$ in the
approximation discussed in Ref.~\cite{Schulz:2012wu}, whose notation
we follow.  For $m,n\ne0$, the derivation is a minor generalization of
that in Ref.~\cite{Schulz:2004tt}, obtained by including all of the
light $\CN=4$ degrees of freedom in the dual toroidal orientifold, and
not just the exact moduli of the $\CN=2$ theory.

We now define the remaining quantities appearing in
Eq.~\eqref{eq:Neq4Metric}.  The 2D metric $g_{\r\s}$, for $\r,\s=1,2$
is an arbitrary $T^2$ metric.  The 3D metric $G_{\a\b}$, for
$\a,\b=1,2,3$, is an arbitrary $T^3$ metric, and $\D = \det^{1/2}(G)$.
The quantity $Z$ satisfies a Poisson equation on this $T^3$,
\begin{equation}
  -\nabla_G^2 Z = \sum_\text{sources $s$}Q_s\bigl(\d^3(\bx-\bx^s)-1\bigr),
\end{equation}
where $s$ runs over $I=1,\dots,M$, $I'=1',\dots,M'$, and $O_i$ for
$i=1,\dots,8$.  Here, (i) $Q_I = 1$ for a source at $\bx^I$; (ii)
$Q_{I'} = 1$ for an image source at $\bx^{I'} = -\bx^I$; and (iii)
$Q_{O_i} = -4$ for sources at the $2^3=8$ $\IZ_2$ fixed points on
$T^3$ where each of the $x^\a$ equals $0$ or $\half$.  We adopt the
convention
\begin{equation}
  \int d^3x Z =1.
\end{equation}
A different convention can be absorbed into a redefinition of $V_4$
and $G_{\a\b}$.  The solution can be expressed as
\begin{equation}
  Z = 1 + \sum_\text{sources $s$}Z_s,
  \quad\text{with}\quad
  Z_s(\bx) = Q_sK(\bx,\bx^s),
\end{equation}
where $K(\bx,\bx')$ is a Green's function on $T^3$ satisfying
\begin{equation}
  \nabla_G^2K(\bx,\bx') = \d^3(\bx-\bx')-1,
  \quad\text{with}\quad
  \int d^3y K(\bx,\bx') = 0.
\end{equation}
The connection $A$ satisfies
\begin{equation}\label{eq:Adefinition}
  dA = \star_G dZ - 2m(\eta^1\w dx^2 -\eta^2\w dx^1),
\end{equation}
where $\star_G$ denotes the Hodge star operator in the metric
$G_{\a\b}$ on $T^3$.  In what follows, it will be convenient to write
\begin{equation}
  \eta^4 = dx^4 + A,
\end{equation}
and to define $A_I$ and $A_{I'}$ satisfying
\begin{equation}
  dA_I = \star_G dZ_I,\quad
  dA_{I'} = \star_G dZ_{I'}.
\end{equation}

%%%%%%%%%%%%%%%%%%%%%%
%%%   4.2. Frame   %%%
%%%%%%%%%%%%%%%%%%%%%%

\subsection{Frame}
\label{sec:Frame}

Let $E^\a{}_\b$ be a vielbein for $G_{\a\b}$ and $e^\r{}_\s$ be a
vielbein for $g_{\r\s}$,
\begin{equation}
G_{\a\b} = \d_{\g\d}E^\g{}_\a E^\d{}_\b,\quad
g_{\r\s} = \d_{\t\u}e^\t{}_\r e^\u{}_\s,.
\end{equation}
The natural coframe for the metric~\eqref{eq:Neq4Metric} realizes a
$\text{6D} \to \text{4D} + \text{2D}$ splitting of the coframe bundle
$\CF^*\CX_{m,n}$ into two subbundles $\CF_4^*\CX_{m,n}$ and
$\CF_2^*\CX_{m,n}$:
\begin{equation}
  \begin{split}
    \text{4D coframe:}\quad & 
    \th^\a = (2V_4)^{1/4}\D^{-1/2}Z^{1/2}E^{-1\,a}{}_\b dx^\b,\quad
    \th^4 = (2V_4)^{1/4}\D^{1/2}Z^{-1/2}\eta^4.\\
    \text{2D coframe:}\quad & 
    w^\r = e^\r{}_\s\eta^\s.
  \end{split}
\end{equation}
The structure group of the 4D subbundle $\CF_4^*\CX_{m,n}$ spanned by
the first line is $\SU(2)$, and the quantities on the second line are
global 1-forms trivializing the 2D subbundle $\CF_2^*\CX_{m,n}$.

%%%%%%%%%%%%%%%%%%%%%%%%%%%%%%%%%%%%%%%%%%%%%%%%%%%%%%%%%%%%%%%%%%%%%%%%%%
%%%   4.3. Spin connection, torsion, and twisted exterior derivative   %%%
%%%%%%%%%%%%%%%%%%%%%%%%%%%%%%%%%%%%%%%%%%%%%%%%%%%%%%%%%%%%%%%%%%%%%%%%%%

\subsection{Spin connection, torsion, and twisted exterior derivative}

The decomposition $\CF^* = \CF_4^*\oplus\CF_2^*$ means that the
connection 1-form $\o^A{}_B$ should decompose into an $\so(4)$-valued
connections 1-form $\o^m{}_n$ and $\so(2)$-valued connection 1-form
$\o^\r{}_\s$. The statement that $\CF_4^*$ has structure group $SU(2)$
and $\CF_2^*$ has trivial structure group means that we should have
$\o^m{}_n\subset\su(2)\subset\so(4)$ and $\o^\r{}_\s = 0$, for the
appropriate choice of torsion.  Therefore, the first of the Cartan
structure equations
\begin{equation}
    d\th^A + \o^A{}_B\th^B = T^A,
\end{equation}
should take the form
\begin{equation}\label{eq:CartanEqs}
    d\th^\a + \o^\a{}_\b\th^b + \o^\a{}_4\w\th^4 = T^\a,\quad
    d\th^4 + \o^4{}_\b\th^b = T^4,\quad
    dw^\r = T^r,
\end{equation}
where $T^A = (T^\a, T^4; T^\r)$ is the torsion 2-form.  We can
alternatively write these equations as
\begin{equation}
  d_T\th^A + \o^A{}_B = 0,
\end{equation}
where
\begin{equation}
  d_T\th^A = d\th^A - T^A,
\end{equation}
is the twisted exterior derivative~\eqref{eq:dtwisted} acting on
1-forms.

Let us define a twisted exterior derivative on $\CX_{m,n}$ by
\begin{equation}
  \begin{split}
    & d_T (dx^\a) = d(dx^\a), \quad\text{for}\quad \a=1,2,3,\\
    & d_T \eta^4  = d\eta^4 + 2m(\eta^1\w dx^2 - \eta^2\w dx^1),\\
    & d_T \eta^1  = d\eta^1 -2n dx^1\w dx^3,\\
    & d_T \eta^2 = d\eta^2 -2n dx^2\w dx^3.
  \end{split}
\end{equation}
Then, if we express the torsion 2-form in terms of the basis
$(dx^\a,\eta^4;\eta^\r)$ (and its dual) rather than
$(\th^\a,\th^4;\th^r)$ (and its dual), we have
\begin{equation}
  \begin{split}
    T^\a &= 0,\\
    T^4  &= -2m(\eta^1\w dx^2 - \eta^2\w dx^1),\\
    T^1  &= 2n dx^1\w dx^3,\\
    T^2  &= 2n dx^2\w dx^3.
  \end{split}
  \qquad\qquad\bigl(\text{superscripts in the }(dx^\a,\eta^4;\eta^\r)\text{ basis}\bigr)
\end{equation}
In Sec.~\ref{sec:CYMetric}, we will interpret the topology of
$\CX_{m,n}$ as that of a $T^4$ fibration in the $y^1,y^2,x^3,x^4$
directions over a $\IP^1\cong T^2/\IZ_2$ in the $x^1x^2$ directions.
The fibration gives a splitting of the cohomology of
$\CX_{m,n}$,\footnote{See the discussion of Leray spectral sequences
  in Ref.~\cite{GandH}.  In general, the fibration gives a filtration
  of the cohomology, but when the fiber is K\"ahler, as is the case
  here, we obtain a splitting.} which decomposes the cohomology group
$H_{p+q}(\CX_{m,n},\IZ)$ into the sum of subgroups
$H_{(T)}^p(\IP^1,R^q\pi_*\IZ)$ of degree $p$ on the $\IP^1$ base and
$q$ on the abelian surface fiber.  The torsion is a map on
differential forms preserving the fiber degree and increasing the base
degree by 1.  It can be interpreted as a map $T\colon
H_{(T)}^p(\IP^1,R^q\pi_*\IZ)\to H_{(T)}^{p+1}(\IP^1,R^q\pi_*\IZ)$.

With these definitions, $\eta^\a$ and $\eta^\r$ are twisted closed,
and $\eta^4$ satisfies
\begin{equation}
  d_T\eta^4 = \star_G dZ.
\end{equation}
Therefore, 
\begin{equation}
  d_T{}^2(dx^\a) =  d_T^2\eta^r = 0
  \quad\text{and}\quad
  d_T{}^2\eta^4 = -\nabla^2_GZ dx^1\w dx^2\w dx^3 \ne0.
\end{equation}
The last equation appears at first glance to contradict the property
$d_T{}^2=0$ needed to construct a twisted cohomology ring from $d_T$.
However, this is not the case, since the 1-forms $dx^\a$ and $\eta^4$
are odd under the the involution~\eqref{eq:Involution}.  A basis of
$\IZ_2$ invariant products is
\begin{equation}
  dx^\a\w\eta^4
  \quad\text{and}\quad
  \half\e_{\a\b\g} dx^\b\w dx^\g,
\end{equation}
which are indeed annihilated by $d_T{}^2$.

With this choice of torsion 2-form, it is possible to solve the Cartan
structure equations~\eqref{eq:CartanEqs} for the spin connection.  We
find
\begin{equation}\label{eq:spinconnection}
  \begin{split}
    \o^\ahat{}_\bhat &= \tfrac12\left(\pd_\bhat\log Z\th^\ahat 
      -\pd^\ahat\log Z\th_\bhat\right)
    +\tfrac12 F_{\hat4\,\bhat}{}^\ahat\th^{\hat4},\\
    \o^{\hat4}{}_{\bhat} &= - \o^\bhat{}_{\hat4} 
    = -\tfrac12\pd_{\bhat}\log Z\th^{\hat4}
    -\tfrac12 F^{\hat4}_{\ahat\bhat}\th^\ahat,
  \end{split}
\end{equation}
with all other components vanishing.  Therefore, the spin connection
indeed decomposes in the way described at the beginning of this
section.  Here, we have included hats to clearly distinguish frame
indices from coordinate indices.  We define $F^{\hat4} =
\th^{\hat4}{}_4 F = (2V_4)^{1/4}\D^{1/2}Z^{-1/2} F$, where $F = dA$.
The other quantities appearing in Eq.~\eqref{eq:spinconnection} are
defined in the standard way.

%%%%%%%%%%%%%%%%%%%%%%%%%%%%%%%%%%%%%%%
%%%   4.4. Twisted harmonic forms   %%%
%%%%%%%%%%%%%%%%%%%%%%%%%%%%%%%%%%%%%%%

\subsection{Twisted harmonic forms}
\label{sec:TwistedHarmForms}

%%%%%%%%%%%%%%%%%%%
%%%   1-forms   %%%
%%%%%%%%%%%%%%%%%%%

\subsubsection*{1-forms}

The $\IZ_2$ invariant 1-forms $\eta^1$ and $\eta^2$ are annihilated by
$d_T$, as are their Hodge duals, which are proportional to
$\eta^2\w\g$ and $\eta^1\w\g$, 
\begin{equation}
  \g = 2 Z dx^1\w dx^2\w dx^3\w \eta^4 = 2 Z dx^1\w dx^2\w dx^3\w dx^4.
\end{equation}
Therefore, $\eta^1$ and $\eta^2$ are twisted harmonic 1-forms on
$\CX_{m,n}$.  The 4-form $\g$ is related to the volume form on
$\CF_4^*$ by
\begin{equation}
  \Vol_{(4)} = \th^1\w\th^2\w\th^3\w\th^4 = V_4\g.
\end{equation}

%%%%%%%%%%%%%%%%%%
%%%  2-forms   %%%
%%%%%%%%%%%%%%%%%%

\subsubsection*{2-forms}

The 4D coframe $\CF^*_4\CX_{m,n}$ admits an almost hypercomplex structure
$(\CJ^\a)_m{}^n$, for $\a=1,2,3$, whose corresponding triple of 2-forms
$(j^\a)_{mn} = (j^\a)_m{}^pg_{pn}$ is
\begin{equation}
  \begin{split}
    j^1 &= \th^1\w\th^4 + \th^2\w\th^3,\\
    j^2 &= \th^2\w\th^4 + \th^3\w\th^1,\\
    j^3 &= \th^3\w\th^4 + \th^1\w\th^2.\\
  \end{split}
\end{equation}
These differential forms are closed with respect to the twisted exterior
derivative $d_T$, and are selfdual with respect to the Hodge star
operator $\star_4$ on $\CF_4^*$, defined from the metric
\begin{equation}
  ds^2_4 = (\th^1)^2 + (\th^2)^2 + (\th^3)^2 + (\th^4)^2.
\end{equation}
Since the 6D Hodge star operator acts on these forms as $\star = w^1\w
w^2\w\star_4$, with $w^1$ and $w^2$ closed, we conclude that both
$j^\a$ and $\star j^\a$ are closed.  That is, the $j^\a$ are twisted
harmonic forms on $\CX_{m,n}$, annihilated by the twisted Laplace-de
Rham operator
\begin{equation}
  \D_T = d_T d_T{}^\dagger + d_T{}^\dagger d_T,
\end{equation}
where $d_T{}^\dagger = \star\,d_T\,\star$.  To obtain the complete
list of twisted harmonic 2-forms, the steps are analogous to those for
the harmonic forms of K3 in the corresponding approximate
metric~\cite{Schulz:2012wu}.  We follow Sec.~3.4 of
Ref.~\cite{Schulz:2012wu} closely in the remainder of this section.

The $j^\a$ can be decomposed as sums of pairs of closed 2-forms
\begin{equation}\label{eq:jvsomega}
  j^\a = \sqrt{\frac{V_4}2}(\o^\a + \o_\a),
\end{equation}
where
\begin{equation}
  \begin{split}
  \o^\a &= 2\bar\th^\a\w\bar\th^4 +
  \Bigl(\frac{Z-1}{Z}\Bigr)\e_{\a\b\g}\bar\th^\b\w\bar\th^\g
  + d_T\bar\l_\a,\\
  \o_\a &= \frac1Z \e_{\a\b\g}\bar\th^\b\w\bar\th^\g-d_T\bar\l_\a.
\end{split}
\end{equation}
Here, we have included a possible twisted exact term in the definition
of $\o^\a$ and $\o_\a$, and bars denote a coframe with respect to a
``unit'' 4D metric $d\bar s_4^2$, defined by $\th^m =
(2V_4)^{1/2}\bar\th^m$.  We choose $\bar\l_\a$ so that $\o^\a$ and
$\o_\a$ are twisted harmonic.  Then $\bar\l_\a$ satisfies
\begin{equation}
  \sqrt{\frac{V_4}{2}}\bigl(d_T\bar\l_\a + \star_4 d_T\bar\l_\a\bigr) 
  = \frac{1-Z}{Z}j^\a.
\end{equation}

Finally, the following anti-selfdual 2-forms can be shown to be
twisted harmonic:
\begin{equation}
  \begin{split}
    \o_I &=  \Bigl(\frac{Z_I-Z_{I'}}{Z}\Bigr)_{,\,\a} 
    \Bigl(dx^\a\w(dx^4+A)-
    \frac{Z}2 G^{\a\a'}\e_{\a'\b\g}dx^\b\w dx^\g\Bigr)\\
    &= -d\Bigl((A_I-A_{I'}) - \frac{(Z_I-Z_{I'})}{Z}(dx^4+A)\Bigr),
    \quad\text{for}\quad
    I=1,\dots,M.
  \end{split}
\end{equation}

In summary, a basis of twisted harmonic 2-forms is
\begin{equation}\label{eq:ApproxHarmBasis}
  \o_a = (\o^\a,\o_\a,\o_I),
  \quad\text{together with}\quad
  \eta^1\w\eta^2.
\end{equation}
It is possible to show by an analogous computation to that in App.~E.4
of Ref.~\cite{Schulz:2012wu} that
\begin{equation}\label{eq:EtaIntegral}
  \int_{\CX_{m,n}}\eta^1\w\eta^2\w\o_a\w\o_b = \eta_{ab},
\end{equation}
where $\eta_{ab}$ takes the block form
\begin{equation}
  \eta =
  \begin{pmatrix}
    0 & 2 & 0\\
    2 & 0 & 0\\
    0 & 0 & -1
  \end{pmatrix}.
\end{equation}

Note that $\o^\a\w\o_\b = 2\g\,\d^\a{}_\b$ and that, from
Eq.~\eqref{eq:EtaIntegral}, $\o_I\w\o_J$ is in the same twisted
cohomology class as $-\g\,\d_{IJ}$.

%%%%%%%%%%%%%%%%%%%%%%%%%%%%%%%
%%%   Higher degree forms   %%%
%%%%%%%%%%%%%%%%%%%%%%%%%%%%%%%

\subsubsection*{Higher degree forms}

The twisted harmonic forms of higher degree are obtained from products
of the twisted harmonic 1-forms and 2-forms listed above.  Thus, we
see that $\Harm(\CX_{m,n}) = \Harm_{(4)}\times\Harm_{(2)}$, where
\begin{equation}
  \Harm_{(4)} = \langle 1,\o_a,\g\rangle
  \quad\text{and}\quad
  \Harm_{(2)} = \langle 1,\eta^\r\rangle.
\end{equation}
Here, angle brackets denote the span of the quantities enclosed.  The
cohomology ring similarly factorizes.

%%%%%%%%%%%%%%%%%%%%%%%%%%%%%%%%%%%%%%%%%%%%%%%%%%%%%%%%%%%%%%%%%%%%%%%%
%%%   4.5. Moduli space of almost hypercomplex structure on CF_4^*   %%%
%%%%%%%%%%%%%%%%%%%%%%%%%%%%%%%%%%%%%%%%%%%%%%%%%%%%%%%%%%%%%%%%%%%%%%%%

\subsection{Moduli space of almost hypercomplex structure on
  $\CF_4^*$}
\label{sec:ModAHS}

The metric~\eqref{eq:Neq4Metric} depends explicitly on moduli $V_4$,
$G_{\a\b}$, $x^{I\a}$, $g_{\r\s}$ and implicitly on three additional
moduli $\b^{\a\b}$, defined as follows.  Since
Eq.~\eqref{eq:Adefinition} determines $A$ only up to an additive shift
by a constant 1-form $\b_\a dx^\a$, we can write
\begin{equation}
  A = A^0 + \b_\a dx^\a,
\end{equation}
where $A^0$ is a fiducial connection independent of $E$ and $\b$, but
depending on $M$ sources locations $\bx^I$ on $T^3$.  It is convenient
to trade $\b_\a$ for a bivector $\b^{\a\b} = \e^{\a\b\g}\b_\g$.

Since the Hodge star operation depends on the metric moduli, so do the
harmonic forms.  It is straightforward to see the that the harmonic
forms of the previous section do not depend on $g_{\r\s}$.  The
$\eta^\r$ are moduli independent, and the $\o_a$ depend on $E$, $\b$,
and $x$.  Let us define moduli independent forms $\xi_a$ by
\begin{equation}\label{eq:ModIndepXi}
  \xi_a = \o_a\Big|_{\displaystyle(E,\b,x)=(1,0,0)}. 
  % \text{ evaluated at } (E,\b,x) = (1,0,0).
\end{equation}
Then, for suitable conventions on $A^0$, one can show that the twisted
cohomology classes of the $\o_a$ and $\xi_a$ are related by
\begin{equation}\label{eq:OmegaVsXi}
  [\o_a] = V(E,\b,x)_a{}^b[\xi_b],
\end{equation}
where 
\begin{equation}\label{eq:SOvielbein}
  \begin{split}
    V(E,\b,x) = V(E)V(\b)V(x)
    &= 
    \begin{pmatrix}
      E & 0       & 0\\
      0 & E^{-1T} & 0\\
      0 & 0       & 1
    \end{pmatrix}
    \begin{pmatrix}
      1 & -\b & -0\\
      0 &  1 & 0\\
      0 &  0 & 1
    \end{pmatrix}
    \begin{pmatrix}
      1 & x^Tx & 2x^T\\
      0 & 1    & 0\\
      0 & x    & 1
    \end{pmatrix}\\
    &=
    \begin{pmatrix}
      E & -E C     & 2Ex^T\\
      0 & E^{-1T} & 0\\
      0 & x       & 1
    \end{pmatrix},
  \end{split}
\end{equation}
with $C^{\a\b} = \b^{\a\b} - \d_{IJ}x^{I\a}x^{J\b}$.  The nontrivial
part of this relation to prove is the $x$ dependence.  Setting $E=1$
and $\b=0$ for simplicity, the identification of the $x$ dependence of
both sides of Eq.~\eqref{eq:OmegaVsXi} is closely analogous to that of
the corresponding K3 discussion in Sec.~3.4 of
Ref.~\cite{Schulz:2012wu}, to which we refer the reader for further
discussion.  The $x^{I\a}$ correspond to D6 positions on $T^3$ in the
D6/O6 dual of the type IIB $T^6/\IZ_2$ orientifold after T-dualizing a
$T^3$ in the $T^6$.  The choice $x^{I\a} = 0$ for $I=1,\dots,M$
describes a metric~\eqref{eq:Neq4Metric} with a curve of $D_M$
singularities, and corresponds to choosing all $M$ D-branes coincident
in the dual orientifold with $\CN=2$ flux, yielding enhanced $SO(2M)$
gauge symmetry.

Comparing Eqs.~\eqref{eq:jvsomega} and \eqref{eq:OmegaVsXi}, we see
that the matrix $V(E,\b,x)$ parametrizes the moduli space of almost
hypercomplex structure on $\CF_4^*$.  Choices of $\o_\a$ differing by
$SO(3)$ rotation of $(\o^\a,\o_\a)$ and $SO(3+M)$ rotation of $\o_I$
gives the same almost hypercomplex structure (AHS).\footnote{Note that
  $\o^\a$ and $\o_\a$ are in the same representation of $\SO(3)$,
  since $O^{-1T}=O$ for $O\in\SO(3)$.}  The moduli matrix $V(E,\b,x)$
is best viewed as a vielbein for the coset
\begin{equation}
  \CM_\text{AHS} = \bigl(SO(3)\times SO(3+M)\bigr)\backslash
  SO(3,3+M)/\G_{3,3+M},
\end{equation}
where $\G_{3,3+M}$ is the discrete group of lattice isomorphisms of
the twisted cohomology lattice of $\CF_4^*\CX_{m,n}$.  The
$\SO(3)\times\SO(3+M)$ acts on the left (frame) index of $V_a{}^b$,
and $\G_{3,3+M}$ acts on the right index of $V_a{}^b$ by
Eq.~\eqref{eq:OmegaVsXi} since it induces an action on $\xi_b$.  The
space of 4D metrics on $\CF_4^*$ is the space $\IR^1_{>0}$ of overall
volume modulus $V_4$ times the space of almost hypercomplex structure
$\CM_\text{AHS}$.

%%%%%%%%%%%%%%%%%%%%%%%%%%%%%%%%%%%%%%%%%%%%%%%%%%%%%%%%%%%%%%%%%%%%%
%%%   4.6. Moduli space of almost complex structure on $\CF_2^*   %%%
%%%%%%%%%%%%%%%%%%%%%%%%%%%%%%%%%%%%%%%%%%%%%%%%%%%%%%%%%%%%%%%%%%%%%

\subsection{Moduli space of almost complex structure on $\CF_2^*$}

An almost complex structure can be defined on $\CF_2^*\CX_{m,n}$ by
taking the forms of type (1,0) and (0,1) to be
\begin{equation}
  w = w^1 + iw^2
  \quad\text{and}\quad
  \bar w = w^1-iw^2.
\end{equation}
Parametrizing the 2D metric on $\CF_2^*$ as
\begin{equation}
  ds_2^2 = g_{\r\s}\eta^\r\eta^\s 
  = \frac{V_2}{\im\t_2} \bigl|\eta^1+\t_2\eta^2|^2,
\end{equation}
this is equivalent to
\begin{equation}
  w = \Bigl(\frac{V_2}{\im\t_2}\Bigr)^{1/2}\eta,
  \quad\text{where}\quad
  \eta = \eta^1+\t_2\eta^2.
\end{equation}
The space of 2D metrics $g_{\a\b}$ is
\begin{equation}
  \SO(2)\backslash\GL(2)/\G_2,
\end{equation}
from the choice of vielbein $w^\r{}_\s\in\GL(2)$, modulo $\SO(2)$
rotation on the left and change of twisted cohomology lattice basis
$\G_2\cong\GL(2,\IZ)$ on the right.  Equivalently, it is the space
$\IR^1_{>0}$ of overall volume modulus $V_2$ times the space
\begin{equation}
  \CM_\text{ACS} = U(1)\backslash\SL(2)/\PSL(2,\IZ)
\end{equation}
of almost complex structure moduli $\t_2$.

%%%%%%%%%%%%%%%%%%%%%%%%%%%%%%%%%%%%%%%%%%%%%%%%%%%%%%%%%%%%%%%%%%%
%%%   4.7. Moduli space of SU(2) structure metrics on X_{m,n}   %%%
%%%%%%%%%%%%%%%%%%%%%%%%%%%%%%%%%%%%%%%%%%%%%%%%%%%%%%%%%%%%%%%%%%%

\subsection{Moduli space of $\SU(2)$ structure metrics on
  $\CX_{m,n}$.}

We have shown that the moduli $V_4$, $G_{\a\b}$, $x^{I\a}$ and
$g_{\a\b}$ of the first-order $\CN=4$ metric~\eqref{eq:Neq4Metric}
span a moduli space
\begin{equation}
  \CM = \IR^1_{>0}\times\IR^1_{>0}
  \times\CM_\text{AHS}\times\CM_\text{ACS}.
\end{equation}

To see that the metric on this moduli space coincides with the coset
metric for the $\CM_\text{AHS}$ and $\CM_\text{ACS}$ factors, we
proceed as follows.  Let us assume that the metric on moduli space
agrees with the moduli space metric from naive dimensional reduction,
discussed in Sec.~3.5.2 of Ref.~\cite{Schulz:2012wu},
\begin{equation}\label{eq:NaiveModMet}
  ds^2_{\CM,\text{Naive}} = \frac23 \Bigl(\frac{\d V_6}{V_6}\Bigr)^2
    + 2 \int_{\CX_{m,n}} d^6x \sqrt{\Gtilde}\Bigl(
      \tfrac14\Gtilde^{mp}\Gtilde^{nq}\d\Gtilde_{mn}\d\Gtilde_{pq}
      - \tfrac14 (\Gtilde^{mn}\d\Gtilde_{mn})^2\Bigr),
\end{equation}
where $V_6 = V_4V_2$.  Here, $\tilde G$ is the 6D unit metric
\begin{equation}
  d\tilde s^2 = (V_6)^{-1/3}\Bigl[(2V_4)^{1/2}d\bar s_4^2 
  + V_2d\bar s_2^2\Bigr],
\end{equation}
in terms of the 4D and 2D unit metrics on $\CF_4^*$ and $\CF_2^*$,
\begin{align}
  d\bar s_4^2 &= \Gbar_{mn}dx^mdx^n = 
  \D^{-1}Z G_{\a\b} dx^\a dx^\b + \D Z^{-1}(dx^4+A)^2,\\
  d\bar s_2^2 &=  \frac1{\im\t_2} \bigl|\eta^1+\t_2\eta^2|^2.
\end{align}
Due to the $\IZ_2$ identification under the
involution~\eqref{eq:Involution}, the 6D and 4D unit metrics have
volume 1/2 in our conventions, and their $\IZ_2$ covering spaces have
volume 1.

By steps closely analogous to those in Ref.~\cite{Schulz:2012wu}, the
naive moduli space metric can be evaluated, giving
\begin{equation*}
  ds^2_{\CM,\text{Naive}} = \frac23 \Bigl(\frac{\d V_6}{V_6}\Bigr)^2
  + \Bigl(\frac{\d(V_4^{1/2}V_6^{-1/3})}{V_4^{1/2}V_6^{-1/3}}\Bigr)^2
  + \Bigl(\frac{\d(V_2V_6^{-1/3})}{(V_2V_6^{-1/3})}\Bigr)^2
  + ds^2_\text{AHS} + ds^2_\text{ACS},
\end{equation*}
which simplifies to
\begin{equation}\label{eq:SU2ModSpaceMetric}
  ds^2_{\CM,\text{Naive}} =
  \frac23\Bigl(\frac{\d V_6}{V_6}\Bigr)^2 +
  \frac13\Bigl(\frac{\d V_\text{rel}}{V_\text{rel}}\Bigr)^2
   + ds^2_\text{AHS} + ds^2_\text{ACS}.
\end{equation}
Here, $V_\text{rel} = V_2/(V_4)^{1/2}$ and the last two
terms are the natural coset metrics on $\CM_\text{AHS}$ and
$\CM_\text{ACS}$,
\begin{subequations}
  \begin{align}
    ds^2_\text{AHS} &= \frac14 G_{\a\g}G_{\b\d}
    \bigl(\d G^{\a\b}d G^{\g\d} + \tilde\d\b^{\a\b}\tilde\d\b^{\g\d}\bigr) 
    + \d_{IJ}G_{\a\b}\d x^{I\a}\d x^{J\b},
    \label{eq:AHSmetric}\\
    ds^2_\text{ACS} &= \frac12\Bigl|\frac{\d\t_2}{\im\t_2}\Bigr|^2,
    \label{eq:ACSmetric}
  \end{align}
\end{subequations}
where
\begin{equation}\label{eq:dtildeb}
  \tilde\d\b^{\a\b} = \d\b^{\a\b} - x^{I\a}\d x^{I\b}+x^{I\b}\d
  x^{I\a}.
\end{equation}

For each choice of metric moduli, we have a unique triple $j^\a$ from
Sec.~\ref{sec:TwistedHarmForms} and unique $w^1,w^2$ from Sec.~\ref{sec:Frame}.
Together, these define a unique $SU(2)$ structure on $\CX_{m,n}$ with
$SU(2)$ invariant tensors~\eqref{eq:SU2InvTensors}.

%%%%%%%%%%%%%%%%%%%%%%%%%%%%%%%%%%%%%%%%%%%%%%%%%%%%%%%%%%%%%
%%%   5. First-order Calabi-Yau metric and moduli space   %%%
%%%%%%%%%%%%%%%%%%%%%%%%%%%%%%%%%%%%%%%%%%%%%%%%%%%%%%%%%%%%%

\section{First-order Calabi-Yau metric and moduli space}
\label{sec:FirstOrderCYMetric}

%%%%%%%%%%%%%%%%%%%%%%%
%%%   5.1. Metric   %%%
%%%%%%%%%%%%%%%%%%%%%%%

\subsection{Metric}
\label{sec:CYMetric}

Below the scale at which $\CN=4$ is spontaneously broken to $\CN=2$,
the tree level type IIB supergravity description of $T^6/\IZ_2$ with
flux~\cite{Kachru:2002he,Schulz:2002eh,Schulz:2004ub} is dual to IIA
compactified on $\CX_{m,n}$ with ``first-order''
metric~\cite{Schulz:2004tt}
\begin{equation}\label{eq:Neq2Metric}
  ds^2_6 =   Z\Bigl(\frac{2s}{\im\t_2}\bigl|dx^1+\t_2 dx^2\bigr|^2 
  + \nbar h\im\t_1 (dx^3)^2\Bigr)
  + Z^{-1}\frac{\nbar h}{\im\t_1} (\eta^4)^2
  + \frac{\mbar h}{\im\t_2}\bigl|\eta^1+\t_2\eta^2\bigr|^2,
\end{equation}
which is a restricted form of the $\CN=4$
metric~\eqref{eq:Neq4Metric}.  Here, $(\mbar,\nbar) = (m,n)/\gcd(m,n)$
and we identify $\re\t_1$ with $-\b^{12}$.  The quantities
$\b^{23},\b^{31}$ are fixed, as we will see below, and are no longer
moduli.  Compared to Eq.~~\eqref{eq:Neq4Metric}, we have
\begin{equation}\label{eq:Neq2fromNeq4}
  \begin{gathered}
    \D^{-1}G_{\a\b} = \frac1{\im\t_2}|dx^1+\t_2 dx^2|^2 
    + \frac{\nbar h}{2s}(dx^3)^2,\\
    2V_4 = (\nbar h)(2s),\quad
    V_2 = \mbar h,\quad
    \D = \sqrt{\frac{\nbar h}{2s}}\frac1{\im\t_1},\quad
    \b^{12}=-\re\t_2.
  \end{gathered}
\end{equation}

Since the curvatures $d\eta^1=dA^1$, $d\eta^2=dA^2$ and $d\eta^4=dA$
all vanish when restricted to constant $x^1,x^2$, this metric
describes a fibration whose generic fibers at constant $x^1,x^2$ are
tori $T^4_{\{y^1,y^2,x^3,x^4\}}$.  Due to the $\IZ_2$
involution~\eqref{eq:Involution}, the base is $\IP^1 =
T^2_{\{x^1,x^2\}}/\IZ_2$.  This is also borne out in the analysis of
Refs.~\cite{Schulz:2004tt} and \cite{Donagi:2008ht}, which show that
$\CX_{m,n}$ is special type of $T^4$ fibration with projective
embeddings, known as an abelian surface fibration over $\IP^1$.  On
the region $Z>0$, the first-order metric~\eqref{eq:Neq2Metric} is a
positive definite Calabi-Yau metric.  It has not only $SU(2)$
structure, but $SU(3)$ holonomy.  The harmonic K\"ahler form and
holomorphic 3-form are
\begin{subequations}
  \begin{align}
    J &= j^3 + w^1\w w^2 = 
    h\bigl(\mbar\eta^1\w\eta^1 + \nbar dx^3\w\eta^4\bigr)
    + 2s Z dx^1\w dx^2,\label{eq:CYKahlerForm}\\
    \O &= i(j^1+ij^2)\w(w^1+iw^2) =
    N (dx^1 + \t_2 dx^2)\w(\eta^4 -\im\t_1Z dx^3)\w(\eta^1+\t_2\eta^2),
    \label{eq:CYHol3Form}
  \end{align}
\end{subequations}
where
\begin{equation}
  N = \sqrt{\frac{2V_6}{\im\t_1(\im\t_2)^2}}.
\end{equation}
It is straightforward to check that $dJ=d\O=0$, and that
\begin{equation}
  *J = \frac12 J\w J,\quad
  *\O = i\bar\O,
\end{equation}
so that $J$ and $\O$ are also co-closed.
The normalization factor $N$ ensures that
\begin{equation}
  V_6 = \frac16\int_{\CX_{m,n}}J\w J\w J = 
  \frac{i}8\int_{\CX_{m,n}}\O\w\bar\O.
\end{equation}
Finally, we write $\O = N\O_\text{hol}$, where $\O_\text{hol}$ depends
on the complex structure moduli purely holomorphically.  The K\"ahler
and complex structure moduli are $h,s,x^{I3}$ and
$\t_1,\t_2,x^{I1}+\t_2 x^{I2}$, respectively, for a total of $2+M$ of
each.  The Hodge numbers $h^{11}=h^{21}=M+2$ can also be deduced from
the number of massless vector and hyper multiplets in the dual
$T^6/\IZ_2$ orientifold.

%%%%%%%%%%%%%%%%%%%%%%%%%%%%%%%%%%%%%%%%%%%%%%%%%%%%%%%%%%%%%%
%%%   5.2. Moduli space of Calabi-Yau metrics on X_{m,n}   %%%
%%%%%%%%%%%%%%%%%%%%%%%%%%%%%%%%%%%%%%%%%%%%%%%%%%%%%%%%%%%%%%

\subsection{Moduli space of Calabi-Yau metrics on $\CX_{m,n}$}
\label{sec:KandCmetrics}

The moduli space metric of the previous section restricts to the
moduli space of the first-order Calabi-Yau metric to give
\begin{equation}\label{eq:Neq2ModMetric}
  ds^2_{\CN=2} = \half\Bigl(\frac{\d V_6}{V_6}\Bigr)^2 +
  2ds^2_\text{K\"ahler} + 2ds^2_\text{Complex},
\end{equation}
where the K\"ahler and complex structure moduli spaces are given by
\begin{subequations}
  \begin{align}
    4ds^2_\text{K\"ahler} &= 2\Bigl(\frac{\d s_2}{s_2}\Bigr)^2
    +\Bigl(\frac{\d s_1}{s_1}\Bigr)^2
    +\frac{2}{s_1s_2}\sum_{I=1}^M\bigl(s_2\d x^{I3}\bigr)^2,
    \label{eq:sKahlerMetric}\\
    4ds^2_\text{Complex} &= 2\Bigl|\frac{\d\t_2}{\t_2}\Bigr|^2
    +\Bigl|\frac{\tilde\d\t_1}{\t_1}\Bigr|^2
    +\frac{2}{\t_1\t_2}\sum_{I=1}^M\bigl|\d x^{I1}+\t_2\d x^{I2}\bigr|^2.
    \label{eq:tauComplexMetric}
  \end{align}
\end{subequations}
Here,
\begin{equation}
  s_1 = 2s,\quad
  s_2 = \nbar h,
  \quad\text{and}\quad
  \tilde\d\t_1 = \d\t_1 + x^{I1}\d x^{I2} - x^{I2}\d x^{I1}.
\end{equation}
To obtain this Calabi-Yau moduli space metric, we substitute the
restricted choice~\eqref{eq:Neq2fromNeq4} into the $\SU(2)$ structure
moduli space metric~\eqref{eq:SU2ModSpaceMetric}, and drop the terms
from $\tilde\d\b^{23}$ and $\tilde\d\b^{31}$.

The first term in Eq.~\eqref{eq:Neq2ModMetric} is as expected.  The
physical moduli space metric for the overall volume modulus from
dimensional reduction differs from the standard Calabi-Yau moduli
space metric by this term.  Note that the moduli space metric of
$\CX_{m,n}$ is self mirror.  The K\"ahler and complex structure moduli
space metrics take the same form up to complexification.  At $\re\t_1
= \re\t_2 = x^{I1} =0$, we have
\begin{equation}
  ds^2_\text{K\"ahler}\leftrightarrow ds^2_\text{Complex}
  \quad\text{under}\quad
  (v_1,v_2,x^{I3})\leftrightarrow (\im\t_1,\im\t_2,x^{I2}).
\end{equation}

The variables appearing in Eq.~\eqref{eq:Neq2ModMetric} are not the
standard K\"ahler and complex structure moduli of $\CX_{m,n}$, but are
related by the nonlinear transformation~\eqref{eq:VSrelations} below.
From Eqs.~\eqref{eq:JOfromSU2}, \eqref{eq:jvsomega}
and~\eqref{eq:OmegaVsXi}, the nontwisted cohomology class of $J$ is
\begin{equation}
  \begin{split}
    [J] &= \half s_2[\xi^3 - C^{3\a}\xi_\a + 2x^{I3}\xi^I 
    +\tfrac{m}{n}\eta^1\w\eta^2] 
    +\half s_1[\xi_3] + s_2,\\
    &= \half s_2[\xi^3+\tfrac{m}{n}\eta^1\w\eta^2 - C^{31}\xi_1-C^{32}\xi_2] 
    +\half\bigl(s_1 + s_2 x^{I3}x^{I3}\bigr)[\xi_3] + s_2x^{I3}[\xi_I],
  \end{split}
\end{equation}
where $C^{3\a} = \b^{3\a} - x^{I3}x^{I\a}$.  Here, the forms $s_2\xi_1
= h d\eta^2$ and $s_1\xi_2 = - h d\eta^1$ are exact.  Dropping exact
terms, we have
\begin{equation}\label{eq:Jclass}
  [J] = \half v^2[\xi^3 +\tfrac{m}{n}\eta^1\w\eta^2] + \half v^1[\xi_3] + v^I[\xi_I],
\end{equation}
where
\begin{equation}\label{eq:VSrelations}
  v^2 = s_2,\quad v^1 = s_1 + s_2 x^{I3}x^{I3},\quad v^I = s_2 x^I.
\end{equation}
This allows us to compute the Calabi-Yau volume
\begin{equation}
  V_6 = \tfrac16\int_{\CX_{m,n}}J\w J\w J = \half s_1s_2{}^2 
  = \tfrac16\tfrac{m}{n}C_{ABC}v^A v^B v^C,
\end{equation}
where
\begin{equation}
  \tfrac16C_{ABC}v^Av^Bv^C = \half v^2(v^1v^2-v^Iv^I).
\end{equation}

Starting from the K\"ahler potential $K_\text{K\"ahler}$,
\begin{equation}\label{eq:KahlerPotv}
  \exp(-K_\text{K\"ahler}) = 8 V_6 = \tfrac14\tfrac{m}{n}C_{ABC}v^Av^Bv^C,
\end{equation}
one can compute the K\"ahler moduli space metric
$\pd_A\bar\pd_BK_\text{K\"ahler}\d v^A\d v^B$ and show via
Eq.~\eqref{eq:VSrelations} that we indeed reproduce the
metric~\eqref{eq:sKahlerMetric}.  This is done in
App.~\ref{app:Kahler}.

A similar statement holds for the complex structure moduli space.
Writing
\begin{equation}\label{eq:TauUrelations}
  \im u^2 = \im\t_2,\quad \im u^1 = \im\t_1 + \im\t_2 x^{I2}x^{I2},
  \quad \im u^I = \im\t_2 x^{2I},
\end{equation}
the complex structure volume is
\begin{equation}
  \begin{split}
    V_\text{cpx} &=
    \tfrac{i}{8}\int_{\CX_{m,n}}\O_\text{hol}\w\bar\O_\text{hol}\\
    &= \half\im\t_1(\im\t_2)^2
    = \tfrac{i}6 C_{ABC}
    \Bigl(\frac{u^A-\bar u^A}2\Bigr)\w
    \Bigl(\frac{u^B-\bar u^B}2\Bigr)\w
    \Bigl(\frac{u^C-\bar u^C}2\Bigr),
  \end{split}
\end{equation}
for the same $C_{ABC}$.  Starting from the K\"ahler potential
$K_\text{cpx}$,
\begin{equation}
  \exp(-K_\text{cpx}) = 8V_\text{cpx}, 
\end{equation}
one can compute the complex structure moduli space metric
$\pd_A\bar\pd_B K_\text{cpx}(u,\bar u)\d u^A\d \bar u^B$ and show via
Eq.~\eqref{eq:TauUrelations} that we also reproduce
Eq.~\eqref{eq:tauComplexMetric}.

The holomorphic 3-form has nontwisted cohomology class
\begin{equation}\label{eq:Hol3Form}
  \O_\text{hol} = \Bigl[\Bigl((\xi^1+\t_2\xi^2)
  +\t_1(\xi_2-\t_2\xi_1)
  + (x^{I1}+\t_2 x^{I2})
  \bigl(2\xi_I+ x^{I1}\xi_1 + x^{I2}\xi_2+ (1+\a)x^{I3}\xi_3\bigr)
  \Bigr)\w(\eta^1+\t_2\eta^2)\Bigr].
\end{equation}
Here, we have fixed the massive $\CN=4$ scalars $\b^{13}$ and $\b^{23}$
to take the values
\begin{equation}
  \b^{13} = \a x^{I1}x^{I3}
  \quad\text{and}\quad
  \b^{23} = \a x^{I2}x^{I3}.
\end{equation}
We now show that $\a=-1$.  Recall that the K\"ahler and complex
structure moduli spaces only locally form a product.  Globally, the
K\"ahler moduli space can be fibered over the complex structure moduli
space, but not vice versa, since the order of logic in constructing a
Calabi-Yau metric is as follows:
\begin{enumerate}
  \item Choose a complex structure on $\CX$.
  \item Choose a real K\"ahler class $[J]\in H^{1,1}(X,\IC)$ of
    positive norm.
  \item By Yau's theorem, there exists a unique Ricci flat metric on
    $X$, with K\"ahler form in this cohomology class.
\end{enumerate}
Therefore, the holomorphic 3-form cannot depend on the K\"ahler
modulus $x^{I3}$, and we conclude that $\a=-1$.  Alternatively, the
form $\xi^3\w(\eta^1+\t_2\eta^2)$ appearing in
Eq.~\eqref{eq:Hol3Form}, with coefficient proportional to $(1+\a)$, is
not closed:
\begin{equation}
  d\bigl(\half\xi^3\w(\eta^1+\t_2\eta^2)\bigr) 
  = m\eta^1\w\eta^2\w(\xi^2-\t_2\xi^3).
\end{equation}
For $d\O_\text{hol}=0$, we require that $\a=-1$.

%%%%%%%%%%%%%%%%%%%%%%%%%%%%%%%
%%%   5.3. Harmonic forms   %%%
%%%%%%%%%%%%%%%%%%%%%%%%%%%%%%%

\subsection{Harmonic forms}
\label{sec:HarmonicForms}

It is convenient to express the quantity $Z$ as
\begin{equation}
  Z = 1 + \Bigl(\frac{\nbar h}{2s}\Bigr)\hat Z,
\end{equation}
where $\hat Z$ satisfies the rescaled Poisson equation
\begin{equation}
  \Bigl(\pd_3{}^2 + \frac{\nbar h\im\t_1}{2s}\nabla_{\hat T^2}^2\Bigr)
  \hat Z = \sum_\text{sources $s$}Q_s\bigl(\d^3(\bx-\bx^I)-1\bigr),\quad
  \int d^3x\hat Z = 0.
\end{equation}
Here $\hat T^2$ is $T^2_{x^1x^2}$ with unit area,
\begin{equation}
  ds^2_{\hat T^2} = \frac{1}{\im\t_2}\bigl|dx^1+\t_2 dx^2\bigr|^2.
\end{equation}

Expressing the K\"ahler form~\eqref{eq:CYKahlerForm} in terms of $\hat
Z$, we have~\cite{Schulz:2004tt}
\begin{equation}\label{eq:JZhat}
  J = s\o_A + h\o_H,
\end{equation}
where
\begin{subequations}\label{eq:omegaHSI}
  \begin{align}
    \o_H &= \mbar \eta^1\w \eta^2 + \nbar
    \hat Z dx^1\w dx^2 + \nbar dx^3\w\eta^4 + \frac{h}{s}d\lambda,
    \label{eq:omegaHSIa}\\
    \o_A &= 2dx^1\w dx^2-d\lambda,\label{eq:omegaHSIb}
  \end{align}
\end{subequations}
and we choose $\lambda$ so that $\o_H$ and $\o_S$ are harmonic.  As in
Sec.~\ref{sec:TwistedHarmForms}, we also have harmonic forms $\o_I$, for
$I=1,\dots,M$.  

In the first-order description~\eqref{eq:Neq4Metric}, $\o_H,\o_S,\o_I$
form a basis for $H^2(\CX_{m,n},\IR)$.  Letting $\CE_I$ denote the
Poincar\'e dual of $\o_I$, we find the following intersection numbers
$A\cdot B\cdot C = \int \o_A\w\o_B\w\o_C$:
\begin{equation}\label{eq:Intersections}
  H^2\cdot A = 2\mbar\nbar,\quad H\cdot \CE_I\cdot
  \CE_J = -\mbar \d_{IJ},\quad\hbox{others} = 0.
\end{equation}

The forms $\o_H,\o_A,\o_I$ are real and moduli dependent.  From
Eq.~\eqref{eq:Jclass}, we have
\begin{equation}
  [\o_H] = [\mbar\eta^1\w\eta^2 + \half\nbar\z^3]
  \quad\text{and}\quad
  [\o_A] = [\z_3],
\end{equation}
and it is convenient to define moduli independent forms
\begin{equation}
  \xi_H = \o_H\Big|_{\displaystyle \b=x=0}
  = \mbar\eta^1\w\eta^2 + \half\nbar\xi^3
  \quad\text{and}\quad
  \xi_A = \o_A\Big|_{\displaystyle \b=x=0}
  = \xi_3,
\end{equation}
with $\xi_I$ defined as in Eq.~\eqref{eq:ModIndepXi}.  In
Sec.~\ref{sec:IntegerHomology}, we will relate the $\xi_a =
(\xi_H,\xi_A,\xi_I)$ to integral classes on $H_4(\CX_{m,n},\IZ)$.  We
denote the dual Poincar\'e dual homology classes by
$(H^0,A^0,\CE_I^0)$.  The form $\eta_H$ restricted to an abelian fiber
$A$ gives the Hodge form $\mbar dy^1\w dy^2 + \nbar dx^3\w dx^4$ of
the abelian surface fiber.  As we will see in
Sec.~\ref{sec:RelationToAHE}, the class $H\in H_4(\CX_{m,n})$ is
closely related to (but not quite the same as) the classes of Hodge
surfaces of the abelian fibration.

%%%%%%%%%%%%%%%%%%%%%%%%%%%%%%%%%%%%%%%%%%%%%%%%%%%%%%%%%%%%
%%%   6. Twisted versus ordinary Calabi-Yau cohomology   %%%
%%%%%%%%%%%%%%%%%%%%%%%%%%%%%%%%%%%%%%%%%%%%%%%%%%%%%%%%%%%%

\section{Twisted versus ordinary Calabi-Yau cohomology}
\label{sec:TwistedVsOrdinary}

%%%%%%%%%%%%%%%%%%%%%%%%%%%%%%%%%%%%%%%%%%%%%%%%%%%%%%%%%%%%%%%%%%
%%%   6.1. Interpretation in terms of supersymmetry breaking   %%%
%%%%%%%%%%%%%%%%%%%%%%%%%%%%%%%%%%%%%%%%%%%%%%%%%%%%%%%%%%%%%%%%%%

\subsection{Interpretation in terms of supersymmetry breaking}

In this section we consider two differential operators on $\CX_{m,n}$:
the ordinary exterior derivative and the twisted exterior derivative
including torsion.  The breaking of $\CN=4$ to $\CN=2$ is conveniently
summarized in the lifting of some of the cohomology classes of the
latter, so that twisted cohomology ring $H^\bullet_\tw(\CX_{m,n},\IZ)$
is larger than the ordinary untwisted cohomology ring
$H^\bullet(\CX_{m,n},\IZ)$.  The former is closely related to the
$\SU(2)$ structure and underlying \mbox{$\CN=4$} supersymmetry, and
the latter to the standard description of a Calabi-Yau 3-fold as a
manifold of $\SU(3)$ holonomy preserving unbroken \mbox{$\CN=2$}
supersymmetry.  Equivalently, some of the zero modes of the twisted
Laplacian on $\CX_{m,n}$ becomes nonzero (massive) modes of the
ordinary Laplacian.\footnote{In fact, the particular breaking of
  $\CN=4$ to $\CN=2$ here is special in that the two gravitino masses
  are equal~\cite{Schulz:2004ub}.  This uniquely determines the form
  of the superhiggs mechanism, and all mass eigenvalues in terms of a
  single gravitino mass.}  

Let us first consider the de Rham cohomology, which includes only the
free part of the integer cohomology and sets to zero the torsion
classes.  (Note that two different meanings of the word ``torsion''
used here.  The distinction should be clear from context.)  From field
counting alone we deduce the lifting
\begin{equation}
  \begin{split}
    \dim\bigl(H_\tw^2(\CX_{m,n},\IR)\bigr) = M+7
    &\quad\to\quad\dim
    \bigl(H^2(\CX_{m,n},\IR)\bigr) = h^{11}= M+2,\\
    \dim\bigl(H_\tw^3(\CX_{m,n},\IR)\bigr) = 2(M+6)
    &\quad\to\quad\dim
    \bigl(H^3(\CX_{m,n},\IR)\bigr) = 2(h^{21}+1) = 2(M+3).
  \end{split}
\end{equation}
In terms of homology, the complete statement is
\begin{equation}\label{eq:ChainBoundaryCycle}
  \begin{split}
    H^\tw_0(\IR) &= \langle\text{1 cycle}\rangle,\\
    H^\tw_1(\IR) &= \langle\text{2 boundaries}\rangle,\\
    H^\tw_2(\IR) &= \langle\text{2 non-closed chains, 3 boundaries, $(M+2)$ cycles}\rangle,\\
    H^\tw_3(\IR) &= \langle\text{3 non-closed chains, 3 boundaries, $2(M+3)$ cycles}\rangle,\\
    H^\tw_4(\IR) &= \langle\text{3 non-closed chains, 3 boundaries, $(M+2)$ cycles}\rangle,\\
    H^\tw_5(\IR) &= \langle\text{2 non-closed chains}\rangle,\\
    H^\tw_6(\IR) &= \langle\text{1 cycle}\rangle.
  \end{split}
\end{equation}
Here, angle brackets mean ``is generated by,'' and the right hand side
is the characterization based on the ordinary non-twisted exterior
derivative.  The word ``boundaries'' refers to non-twisted cycles that
are trivial in real homology; in integer homology, they are the
torsion cycles of Eqs.~\eqref{eq:CompleteIntHomRing}
and~\eqref{eq:TorsionHom}.

%%%%%%%%%%%%%%%%%%%%%%%%%%%%%%%%%%%
%%%   6.2. Twisted cohomology   %%%
%%%%%%%%%%%%%%%%%%%%%%%%%%%%%%%%%%%

\subsection{Twisted cohomology}

A convenient of basis of global 1-forms and 2-forms that generate the
twisted cohomology ring consists of
\begin{equation}
  \text{1-forms \ } \eta^1,\ \eta^2
  \quad\text{and}\quad
  \text{2-forms \ } \xi^m,\ \xi_m,\ \xi_I
\end{equation} 
of Secs.~\ref{sec:TwistedHarmForms} and \ref{sec:ModAHS}, where
$m=1,2,3,$ and $I = 1,\dots,M$.  (Recall that $M = 16-4mn$.)  Here, the
existence of $\eta^1$ and $\eta^2$ is implied by the $\SU(2)$
structure, and the 2-forms $\xi_a$ arise from deformations of the
triple of $\SU(2)$ invariant 2-forms.  The $\xi_a$ satisfy
\begin{equation}
  \xi^1\w\xi_1 = \xi^2\w\xi_2 = \xi^3\w\xi_3 \equiv 2\g
  \quad\text{and}\quad
  \xi_I\w\xi_J \cong -\g\d_{IJ}\quad\text{in cohomology,}
\end{equation}
with other products of 2-forms vanishing.  Upon integrating, we have
$\int \eta^1\w \eta^2 \w\g = 1$.  Since the generators $\eta^\r$ and
$\xi_a$ are twisted-closed, their products generate the twisted
cohomology ring:
\begin{equation}
  \begin{split}
    H^0_\tw(\IR) &= \langle 1\rangle\quad (\dim{} = 1),\\
    H^1_\tw(\IR) &= \langle \eta^1,\eta^2\rangle\quad (\dim{} = 2),\\
    H^2_\tw(\IR) &= \langle \eta^1\w \eta^2,\xi^m,\xi_m,\xi_I\rangle\quad (\dim{} = M+7),\\
    H^3_\tw(\IR) &= \langle \eta^1\w\xi^m, \eta^1\w\xi_m, \eta^1\w\xi_I; \eta^2\w\xi_m, \eta^2\w\xi^m, \eta^2\w\xi_I\rangle
    \quad (\dim{} = 2(M+6)),\\
    H^4_\tw(\IR) &= \langle \g,\eta^1\w \eta^2\w\xi_m,\eta^1\w \eta^2\w\xi^m,\eta^1\w \eta^2\w\xi_I\rangle
    \quad (\dim{} = M+7),\\
    H^5_\tw(\IR) &= \langle \g\w \eta^2,\g\w \eta^1\rangle\quad (\dim{} = 2)\\
    H^6_\tw \IR) &= \langle \eta^1\w \eta^2\w\g\rangle\quad (\dim{} = 1).
  \end{split}
\end{equation}
These forms have explicit expressions in the first-order metric
(cf.~Secs.~\ref{sec:SU2Metric} and ~\ref{sec:TwistedHarmForms}).

%%%%%%%%%%%%%%%%%%%%%%%%%%%%%%%%%%%%%%%%%%%%%%%%%%%%%%%%%%%
%%%   6.3. Recovery of standard Calabi-Yau cohomology   %%%
%%%%%%%%%%%%%%%%%%%%%%%%%%%%%%%%%%%%%%%%%%%%%%%%%%%%%%%%%%%

\subsection{Recovery of standard Calabi-Yau cohomology}

To see which of these forms is lifted in ordinary (non twisted)
cohomology, we need only the closure conditions on the generating
1-forms and 2-forms.  These are:
\begin{equation}
  \begin{gathered}
    d\eta^1 = -n\xi_2,\quad d\eta^2 = n\xi_1\\
    \half d\xi^1 = -m\eta^1\w\xi_3,\quad \half d\xi^2 = -m \eta^2\w\xi_3,\quad
    \half d\xi^3 = m(\eta^1\w\xi_1+ \eta^2\w\xi_2),
  \end{gathered}
\end{equation}
with $d\xi_m = d\xi_I = 0$.  Note that the first line implies
\begin{equation}
  d(\eta^1\w \eta^2) = -n(\eta^1\w\xi_1 + \eta^2\w\xi_2),
\end{equation}
so that the form $\eta^1\w\xi_1 + \eta^2\w\xi_2$ multiplied by either
$m$ or $n$ is trivial in integer cohomology.  Therefore, the same form
multiplied by $\gcd(m,n)$ is trivial in integer cohomology, and
\hbox{$\eta^1\w\xi_1 + \eta^2\w\xi_2$} generates a $\IZ_{\gcd(m,n)}$
torsion class.  Note also that the linear combination $\xi_H =
\half\nbar\xi^3 + \mbar \eta^1\w \eta^2$ is closed, where
$(\mbar,\nbar) = \frac1{\gcd(m,n)}(m,n)$.

For the closure conditions on 3-forms, the nontrivial equations are
\begin{equation}
  \begin{split}
    &\half d(\eta^2\w\xi^1) = (n\g - m\eta^1\w \eta^2\w\xi_3) 
    = \gcd(m,n)(\nbar\g - \mbar\eta^1\w \eta^2\w\xi_3) ,\\
    &\half d(\eta^1\w\xi^2) = -(n\g - m\eta^1\w \eta^2\w\xi_3) 
    = -\gcd(m,n)(\nbar\g - \mbar\eta^1\w \eta^2\w\xi_3) ,\\
    &\half d(\eta^1\w\xi^3) = -m\eta^1\w \eta^2\w\xi_2,\\
    &\half d(\eta^2\w\xi^3) = m\eta^1\w \eta^2\w\xi_1,
  \end{split}
\end{equation}
with $d(\eta^1\w\xi_m) = d(\eta^2\w\xi_m) = 0$.

For the closure conditions on 4-forms, the nontrivial equations are.
\begin{equation}
  \begin{split}
    \half d(\eta^1\w \eta^2\w\xi^1) = -n \eta^1\w\g,\\
    \half d(\eta^1\w \eta^2\w\xi^2) = -n \eta^3\w\g.\\
  \end{split}
\end{equation}

In the equations above with a $\half$ on the left hand side, the
2-form being differentiated is not integral, but differs from an
integral form by a closed form.  Therefore, the right hand side is
indeed exact in integer cohomology.  In summary, the free non-twisted
cohomology groups are
\begin{equation}\label{eq:ResultHfree}
  \begin{split}
    H^0_\free &= \langle 1\rangle
    = \IZ,\\
    H^1_\free &= \emptyset,\\ %\varnothing
    H^2_\free &= \langle\xi_3,\xi_H,\xi_I\rangle
    = \IZ^{M+2},\\
    H^3_\free &= \langle w^\r\w\xi_1,w^\r\w\xi_2,w^1\w\xi^1,w^2\w\xi^2,
    w^1\w\xi^2+w^2\w\xi^1,w^\r\w\xi_I\rangle
    /\langle w^1\w\xi_1 + w^2\w\xi_2\rangle
    = \IZ^{2(M+3)},\\
    H^4_\free &= \langle w^1\w w^2\w\xi_H,w^1\w w^2\w \xi_3,\g,w^1\w w^2\w\xi_I\rangle
    /\langle\nbar\g-\mbar w^1\w w^2\w\xi_3\rangle
    = \IZ^{M+2},\\
    H^5_\free &= \emptyset,\\
    H^6_\free &= \langle \eta^1\w \eta^2\w\g\rangle
    = \IZ,
  \end{split}
\end{equation}
and
\begin{equation}\label{eq:ResultHtor}
  \begin{split}
    H^0_\tor &= \emptyset,\\ %\varnothing
    H^1_\tor &= \emptyset,\\ %\varnothing
    H^2_\tor &= \langle \xi_1,\xi_2,\rangle
    = \IZ_n\times\IZ_n,\\
    H^3_\tor &= \langle \eta^1\w\xi_3, \eta^2\w\xi_3, \eta^1\w\xi_1 + \eta^2\w\xi_2\rangle
    = \IZ_m\times\IZ_m\times\IZ_{\gcd(m,n)},\\
    H^4_\tor &= \langle \eta^1\w \eta^2\w\xi_1,\eta^1\w \eta^2\w\xi_2,\nbar\g-\mbar \eta^1\w \eta^2\w\xi_3\rangle 
    = \IZ_m\times\IZ_m\times\IZ_{\gcd(m,n)},\\
    H^5_\tor &= \langle \eta^1\w\g,\eta^2\w\g\rangle
    = \IZ_n\times\IZ_n,\\
    H^6_\tor &= \emptyset.   
  \end{split}
\end{equation}
up to a caveat explained in the discussion of integer cohomology the
next section.

Here is another way to summarize the lifting of forms in going from
twisted to ordinary cohomology:\medskip

\noindent
\textbf{1-forms:} $\eta^1$, $\eta^2$ no longer closed.\medskip

\noindent
\textbf{2-forms:} $n\xi_1$, $n\xi_2$ now exact,\\
\phantom{\textbf{2-forms:}} $\xi^1$, $\xi^2$, one linear combination of $\xi^3$ and $\eta^1\w \eta^2$ 
no longer closed.\medskip

\noindent
\textbf{3-forms:} $m\eta^1\w\xi_3$, $m\eta^2\w\xi_3$, $\gcd(m,n)(\eta^1\w\xi_1+\eta^2\w\xi_2)$ now exact,\\
\phantom{\textbf{3-forms:}} $\eta^1\w\xi^3$, $\eta^2\w\xi^3$, one linear combination of
$\eta^1\w\xi^1$ and $\eta^2\w\xi^2$ no longer closed.\medskip

\noindent
\textbf{4-forms:} $m\eta^1\w \eta^2\w\xi_1$, $m\eta^1\w \eta^2\w\xi_2$, $\gcd(m,n)(\nbar\g-\mbar \eta^1\w \eta^2\w\xi_3)$ now exact,\\
\phantom{\textbf{4-forms:}} $\eta^1\w \eta^2\w\xi^1$, $\eta^1\w \eta^2\w\xi^2$ no longer closed.\medskip

\noindent
\textbf{5-forms:} $n\eta^1\w\g$, $n\eta^2\w\g$ now exact.\medskip

\noindent This supplies the non-closed chains and boundary cycles
required by Eq.~\eqref{eq:ChainBoundaryCycle}.

%%%%%%%%%%%%%%%%%%%%%%%%%%%%%%%%%%%%%%%%%%%%%%
%%%   6.4. The integer homology of X_{m,n} %%%
%%%%%%%%%%%%%%%%%%%%%%%%%%%%%%%%%%%%%%%%%%%%%%

\subsection{The integer homology of $\CX_{m,n}$}
\label{sec:IntegerHomology}

In this section we discuss the integer homology ring of the Calabi-Yau
threefold $\CX_{m,n}$ and its relation to the homology basis
$S^0,H^0,\CE^0_I$ of Eq.~\eqref{eq:AHEintersections}.

%%%%%%%%%%%%%%%%%%%%%%%%%%%%%%%%%%%%%%%%%%%%
%%%   6.4.1. Hodge surfaces of X_{m,n}   %%%
%%%%%%%%%%%%%%%%%%%%%%%%%%%%%%%%%%%%%%%%%%%X

\subsubsection{Hodge surfaces of $\CX_{m,n}$}
\label{sec:HodgeSurfaces}

%%%%%%%%%%%%%%%%%%%%%%%%%%%%%%%%%%%%%%%%%%%%%
%%%   Abelian surfaces and Hodge curves   %%%
%%%%%%%%%%%%%%%%%%%%%%%%%%%%%%%%%%%%%%%%%%%%%

\subsubsection*{Abelian surfaces and Hodge curves}

An abelian surface is a projective variety that has the additional
structure of the abelian group $U(1)^4$.  That is,
\begin{equation}
  A\cong\IC^4/\Lambda\cong T^4,
\end{equation}
with addition of points defined as in $\IC^4$.  For $A$ to be
embeddable in a projective space, it must have K\"ahler class
proportional to an integer Hodge form~\cite{GandH}.  If we choose
coordinates $y^1,y^2,x^3,x^4$ on $A$ with periodicity $y^\r\cong
y^\r+1$ and $x^\r\cong x^\r+1$, then an integer form $m dy^1\w dy^2 +
n dx^3\w dx^4$ is proportional to
\begin{equation}
  \o = \mbar dy^1\w dy^2 + \nbar dx^3\w dx^4,
\end{equation}
where bars denote division by $\gcd(m,n)$.  For an abelian surface, at
most one of $\mbar$ and $\nbar$ can differ from one.  By Poincar\'e
duality, we can view $[\o]$ as an element of either $H^2(A,\IZ)$ or
$H_2(A,\IZ)$, and we will use the same notation for both.  If $[\o]$
is represented by an irreducible Hodge curve $C\subset A$, then, since
$c_1(A)=0$, the genus of this curve is determined by its Euler
characteristic
\begin{equation}\label{eq:mnEuler}
2g-2 = C\cdot C = \int\o\w\o = 2\mbar\nbar\int_{T^4} dy^1\w dy^2\w
dx^3\w dx^4 = 2\mbar\nbar,
\end{equation}
to be $g= \mbar\nbar+1$.

%%%%%%%%%%%%%%%%%%%%%%%%%%%%%%%%%%%%%%%%%%%%%%%%%%%%%%%%%
%%%   Abelian surface fibrations and Hodge surfaces   %%%
%%%%%%%%%%%%%%%%%%%%%%%%%%%%%%%%%%%%%%%%%%%%%%%%%%%%%%%%%

\subsubsection*{Abelian surface fibrations and Hodge surfaces}

To say that the Calabi-Yau manifold $\CX_{m,n}$ is an abelian surface
fibration over $\IP^1$ means that all of the structure of the
preceding paragraph is fibered over $\IP^1$.  A Hodge class $[\o]\in
H^2(\CX_{m,n},\IZ)\cong H_4(\CX_{m,n},\IZ)$ is represented by a Hodge
surface $S$, which is itself fibered by genus $g=\mbar\nbar$ curves
$C$ over $\IP^1$.

Given an embedding $\iota\colon\ S\hookrightarrow\CX_{m,n}$ of the
type we have just described, the projection
$\pi\colon\CX_{m,n}\to\IP^1$ induces a projection
$\pi'=\pi\circ\iota\colon S\to\IP^1$ with base $\pi'(S) = \IP^1$ and
generic fiber $\pi'^{-1}(p)\cong C$ for $p\in\IP^1$.  Then, every
section $\ell$ of $S$ determines a section $\s = \iota(\ell)$ of
$\CX_{m,n}$, since $\ell\subset S$ meets each fiber $A$ in a single
point $p\in C\subset A$.

An abelian surface fibration comes with a zero section, and with the
operation of addition of sections defined.  Given the zero section
$\s_0$, another section $\s$ of $\CX_{m,n}$, and an embedding
$S\hookrightarrow\CX_{m,n}$, we can define a second embedding by
shifting abelian fiber coordinates by $\s-\s_0$.

%%%%%%%%%%%%%%%%%%%%%%%%%%%%%%%%%%%%%%%%%%%%
%%%   The Calabi-Yau threefold X_{m,n}   %%%
%%%%%%%%%%%%%%%%%%%%%%%%%%%%%%%%%%%%%%%%%%%%

\subsubsection*{The Calabi-Yau threefold $\CX_{m,n}$.}

Recall from above that $\CX_{m,n}$ has $b_4=M+2$, where $M=16-4mn$,
with possible values $12, 8, 4, 0$.  The $M+2$ homology classes can be
realized as follows.

For $M\ne0$, each Hodge surface $S$ has $2M$ sections
$\ell_I,\ell'_I$, for $I=1,\dots,M$, with homology classes satisfying
$[\ell_I] + [\ell_I'] = [C']\in H_2(S,\IZ)$, independent of $I$.
Given an embedding $\iota\colon\ S\hookrightarrow\CX_{m,n}$, the $2M$
sections $\ell_I$ and $\ell'_I$ of $S$ determine $2M$ embeddings
$\iota_I$ and $\iota'_I\colon\ S\hookrightarrow\CX_{m,n}$.  One of
these coincides with the original embedding, provided that one of the
sections of $\iota(S)$, denoted $S_0$ coincides with the zero section
$\s_0$ of $\CX_{m,n}$.  We focus on this case, and write $S_I =
\iota_I(S)$ and $S'_I = \iota'_I(S)$, with homology classes satisfying
$[S_I]+[S'_I]=[D]\in H_4(\CX_{m,n},\IZ)$, independent of $I$.  Then,
$H_4^\free(\CX_{m,n},\IZ)$ is generated by the $M+1$ independent classes
from the Hodge surfaces, together with the class of the Abelian
surface fiber $[A]$.

For $M=0$, each Hodge surface $S$ has a single section.  Given an
embedding $\iota\colon\ S\hookrightarrow\CX_{m,n}$, this section of
$S$ determines another embedding $\iota_0\colon\
S\hookrightarrow\CX_{m,n}$.  The two embeddings coincide, provided the
zero section of $\iota(S)$ coincides with the the zero section $\s_0$
of $\CX_{m,n}$.  Again, we focus on this case, and write $S_0 =
\iota(S)$.  Then $H_4^\free(\CX_{m,n},\IZ)$ is generated by $[S_0]$ together
with the class of the Abelian surface fiber $[A]$.

%%%%%%%%%%%%%%%%%%%%%%%%%%%%%%%%%%%%%%%%%%%%%%%%
%%%   6.4.2. The Mordell-Weill lattice D_M   %%%
%%%%%%%%%%%%%%%%%%%%%%%%%%%%%%%%%%%%%%%%%%%%%%%%

\subsubsection{The Mordell-Weil lattice $D_M$}

The sections $\s$ of the abelian surface fibration $\CX_{m,n}$ form a
lattice known as the Mordell-Weil lattice
$\MW(\CX_{m,n})/\MW^\tor(\CX_{m,n})$.  From the group addition law, we
already know that we can add any two sections to obtain a new section.
To define a lattice, we also need an inner product between sections,
which in this context is known as a ``height pairing.''  Since the
sections are curves, they do not generically intersect in the
threefold $\CX_{m,n}$, but do in a surface, and this is exactly the
additional structure we have at our disposal.  We can compute
intersections in a Hodge surface of the abelian fibration.

As discussed above, a Hodge surface $S$ of $\CX_{m,n}$ is a fibered
over $\IP^1$ by curves $C$ of genus $g=\mbar\nbar$, and has $2M$
sections $\ell_I$ and $\ell'_I$, for $I=1,\dots,M$, satisfying
$\ell_I+\ell'_I = C'$, with $C'$ independent of $I$.  The
intersections of these curves in $S$ are
\begin{equation}
  \ell_I{}^2 = \ell'_I{}^2 = -\mbar,\quad
  \ell_I\cdot\ell'_I = \mbar,\quad\text{and}\quad
  \ell_I\cdot\ell_J = \ell'_I\cdot\ell_J = \ell'_I\cdot\ell'_J = 0
  \quad\text{for}\quad I\ne J,
\end{equation}
together with $\ell\cdot C = 1$ and $C^2 = 2\mbar\nbar$.  We assume
that we are given an embedding $\iota\colon\
S\hookrightarrow\CX_{m,n}$ such that one of the sections, denoted
$\ell_0$, maps to the zero section of $\CX_{m,n}$: $\iota(\ell_0) =
\s_0$.  Then, $H_2(S,\IZ)$ is spanned by the $M+1$ independent
sections, together with the genus $\mbar\nbar$ curve $C$, and their
images under $\iota$ likewise span $H_2(\CX_{m,n},\IZ)$ up to torsion
classes.  Thus, for any two sections of $\CX_{m,n}$, we can compute
the intersection of the corresponding homology classes in $S$.  A
height pairing on sections of $\CX_{m,n}$ is given by the intersection
pairing on the orthogonal complement $\langle\ell,C\rangle^\perp$ in
$S$.

On $S$, let us choose $\ell_0 = \ell'_M$.  Then, the projection to
$\langle\ell,C\rangle^\perp$ in $H_2(S,\IZ)$ maps the sections of $S$
as
\begin{equation}
  \begin{split}
    \ell_I & \mapsto\ell_I^\perp = \ell_I-\ell'_M - C,\\
    \ell'_I & \mapsto\ell_I'^\perp = \ell'_I-\ell'_M - C,
    \quad I=1,\dots M,\\
    \ell_M & \mapsto\ell_M^\perp = \ell_M-\ell'_M - 2C,\\
    \ell'_M & \mapsto\ell_M'^\perp = 0.
  \end{split}
\end{equation}
The lattice $\langle\ell,C\rangle^\perp$ is $-\mbar(D_M)$, where
$(D_M)$ denotes the root lattice of $D_M$ the prefactor denotes
$-\mbar$ times the usual Cartan inner product.  If we choose $D_M$
roots $a_I = [\ell_I'^\perp - \ell_{I+1}'^\perp]$ for $I=1,\dots, M-1$
and $a_M = -[\ell_{M-1}^\perp]$, then we have
\begin{equation}\label{eq:DMroots}
  \begin{split}
    a_I &= [\ell'_I - \ell'_{I+1}],\\
    a_{M-1} &= [\ell'_{M-1}-\ell'_M - C],\\
    a_M &= [\ell'_{M-1} - \ell_M + C]
    \quad\text{(using $[-\ell_{M-1}+\ell'_M] = [\ell'_{M-1}-\ell_M]$).}
  \end{split}
\end{equation}
It is straightforward to check that the intersections of
$a_1,\dots,a_M$ as defined by Eq.~\eqref{eq:DMroots} give $-\mbar$
times those of the $D_M$ Dynkin diagram.

%%%%%%%%%%%%%%%%%%%%%%%%%%%%%%%%%%%%%%%%%%%%%%%%%%%%%%%%
%%%   6.4.3. The relation to the basis of Sec. 5.3   %%%
%%%%%%%%%%%%%%%%%%%%%%%%%%%%%%%%%%%%%%%%%%%%%%%%%%%%%%%%

\subsubsection{The relation to the basis of Sec.~\ref{sec:HarmonicForms}}
\label{sec:RelationToAHE}

We would now like to relate the cohomology classes of the Hodge
surfaces $S_I,S'_I$ and Abelian fiber $A$ to those of the basis
$A,H^0,\CE^0_I$ defined in Sec.~\ref{sec:HarmonicForms}.  From
Eq.~\eqref{eq:Intersections}, we have
\begin{equation}
  H\cdot\CE^0_I\cdot\CE^0_J = -\mbar\d_{IJ}.
\end{equation}
Therefore, the curves $e_I = H\cdot\CE^0_I$ form an orthogonal basis
on the surface $H$ with normalization $e_I\cdot_H e_I=-\mbar$, in
terms of which the roots generating a $D_M$ lattice can be realized in
the standard way,
\begin{equation}
  e_1-e_2,\quad e_2-e_3,\quad\dots,\quad e_{M-1}-e_M,\quad
  e_{M-1}+e_M.
\end{equation} 
The only caveat is that $H$ might not be a Hodge surface or even an
integral cycle, and the ``roots'' constructed in this way might not be
in $H_2(\CX_{m,n},\IZ)$.  In fact, this will turn out to be the case.
Nevertheless, it will be helpful to proceed in this way.

Identifying $S$ with $S_0=S'_M$, and using the intersections discussed
in App.~\ref{app:Intersections} below, Eq.~\eqref{eq:DMroots} becomes
\begin{equation}
  \begin{split}
    a_I &= [\ell'_I + \ell'_M] - [\ell'_{I+1}+\ell'_M] =
    \frac1{\mbar\nbar}[S_I-S_{I+1}]\cdot [S],\\
    a_{M-1} &= [\ell'_{M-1} + \ell'_M] - [2\ell'_M+C] =
    \frac1{\mbar\nbar}[S_{M-1}-S_M]\cdot [S],\\
    a_I &= [\ell'_{M-1} +\ell'_M] - [\ell_M+\ell'_M - C] =
    \frac1{\mbar\nbar}[S_{M-1}-S'_M]\cdot [S],
  \end{split}
\end{equation}
or equivalently, using $[S_I+S_I']=[D]$, independent of $I$,
\begin{equation}
  \begin{split}
    a_I &= \frac1{\mbar\nbar}[\half(S_I-S'_I)-\half(S_{I+1}-S'_{I+1})]\cdot [S],\\
    a_{M-1} &= \frac1{\mbar\nbar}[\half(S_{M-1}-S'_{M-1})-\half(S_M-S'_M)]\cdot [S],\\
    a_M &= \frac1{\mbar\nbar}[\half(S_{M-1}-S'_{M-1})+\half(S_M-S'_M)]\cdot [S],\\
  \end{split}
\end{equation}
so that
\begin{equation}
  \begin{split}
    a_I &= [\CE^0_I - \CE^0_{I+1}]\cdot [S],\\
    a_{M-1} &= [\CE^0_{M-1} - \CE^0_M]\cdot [S],\\
    a_M &= [\CE^0_{M-1} + \CE^0_M]\cdot [S]
  \end{split}
\end{equation}
for
\begin{equation}\label{eq:Esoln}
  [\CE^0_I] = \frac1{2\mbar\nbar}[S_I-S'_I].
\end{equation}
If $H^0$ coincides with a Hodge surface, then our identification of
$H^0,\CE^0_I$ is complete.  In fact, $H^0$ does not coincide with a
Hodge surface, but this line of reasoning will lead us to the correct
identification.

To determine $H$ let us start with the two defining properties
$H^0\cdot H^0\cdot H^0 = 0$ and $H^0\cdot H^0\cdot A = 2\mbar\nbar$.
The $\CE^0_I$ as defined in Eq.~\eqref{eq:Esoln}, together with $A$
and any one $S_J$, form a basis for $H_4^\free(\CX_{m,n},\IZ)$. Of this
basis, only $S_J$ has nonzero intersection with $A$.  From
App.~\ref{app:Intersections}, we have
\begin{equation}
  S_J\cdot S_J\cdot A = 2\mbar\nbar.
\end{equation}
Therefore, whether or not the $\CE^0_I$ of Eq.~\eqref{eq:Esoln} are
the correct identification of $\CE^O_I$, $H^0$ has an expansion
\begin{equation}
  [H] = [S_J] + c^a [A] + c^I[\CE^0_I],
\end{equation}
using the $\CE^0_I$ so defined.  By requiring that $H^3 = 0$, we find
from the intersections of App.~\ref{app:Intersections} a family of
solutions for $H^0$ given by
\begin{equation}\label{eq:HwithShifts}
  [H^0] = [S_J] + \frac23\mbar^2\nbar [A] +\text{shifts,}
\end{equation}
where the ``$+$ shifts'' denotes shifts by an arbitrary linear
combination of $[\mbar^2\nbar^2\CE^0_I + \half\mbar^2\nbar A]$ for
$I=1\dots,M$. In the case $M=0$, there is a single $S_0$, and none of
these shifts exist.  Therefore,
\begin{equation}\label{eq:HforMeq0}
  [H^0] = [S_0] + \frac23\mbar^2\nbar [A],
  \quad\text{for $M=0$,}
\end{equation}
which is the case $(m,n) = (\mbar,\nbar) = (4,1)$ or $(1,4)$.  Next,
the from $\CE^0_I\cdot A = 0$, the $\CE^0_I$ must either be as defined
in Eq.~\eqref{eq:Esoln}, or must be a linear combination of the same
classes.  Either way, the condition $H^0\cdot H^0\cdot\CE^0_I = 0$
follows, for $\CE^0_I$ as defined in Eq.~\eqref{eq:Esoln}.  For
$M\ne0$, this uniquely fixes the shifts in Eq.~\eqref{eq:HwithShifts}.
We find
\begin{equation}\label{eq:HforMnonzero}
  [H] = \half[S_J+S'_J] + \frac{\mbar^2\nbar}{6} [A]
  \quad\text{for $M\ne0$,}
\end{equation}
where the class $[S_J+S'_J] = [D]$ is independent of $J$.
Eqs.~\eqref{eq:Esoln}, \eqref{eq:HforMeq0} and \eqref{eq:HforMnonzero}
complete our identification of the basis $[A],[H^0],[\CE^0_I]$ of
Sec.~\ref{sec:HarmonicForms} in terms of the $[A]$ and the Hodge
surfaces.

Thus, $[H^0]$ and $[\CE^0]$ are not actually integer classes, but
rather
\begin{equation}
  \begin{gathered}
    \,[H^0] - \frac{\mbar^2\nbar}{6}[A]\in H_4(\CX_{m,n},\half\IZ)
    \qquad\text{($M$ arbitrary)},\\
    \mbar\nbar[\CE^0_I]\in H_4(\CX_{m,n},\half\IZ)
    \quad\text{and}\quad
    [H^0]\pm\mbar\nbar[\CE^0_I]\in H_4(\CX_{m,n},\IZ)
    \qquad\text{($M$ nonzero).}
  \end{gathered}
\end{equation}
Therefore, in the entries for $H_2^\free(\CX_{m,n},\IZ)$ and
$H_4^\free(\CX_{m,n},\IZ)$, \eqref{eq:ResultHfree}, $\xi_H$ and
$\xi_I$ of should be replaced by the linear combinations
$\xi_H\pm\mbar\nbar\xi_I$ and $\mbar\nbar(\xi_I\pm\xi_I)$ for $M\ne0$,
and $\xi_H$ should be replaced by $2\xi_H$ for $M=0$.

%%%%%%%%%%%%%%%%%%%%%%%%%%
%%%   7. Conclusions   %%%
%%%%%%%%%%%%%%%%%%%%%%%%%%

\section{Conclusions}
\label{sec:Conclusions}

We have shown that the manifolds in class of abelian surface fibered
Calabi-Yau threefolds $\CX_{m,n}$ have a novel interpretation as
manifolds of $\SU(2)$ structure in addition to the conventional
interpretation as manifolds of $SU(3)$ holonomy.  The threefolds are
obtained from duality to the type IIB $T^6/\IZ_2$ orientifold, in
which the choice of flux spontaneously breaks $\CN=4$ to $\CN=2$
supersymmetry.  In the $\SU(2)$ structure description, the topology of
the Calabi-Yau topology spontaneously break $\CN=4$ to $\CN=2$
supersymmetry at a scale that can be made hierarchically larger than
the compactification scale by taking the base to be large.

As a manifestation of the $\SU(2)$ structure, the frame bundle of
$\CX_{m,n}$ splits as the sum of 4D and 2D subbundles, and the moduli
space of $\SU(2)$ structure metrics enlarge the Calabi-Yau metric
moduli space, with a natural interpretation in terms of spaces of
almost hypercomplex structure and almost complex structure on the 4D
and 2D bundles, respectively.  The analysis was facilitated by the
existence of an explicit family of metrics approximating the exact
$\SU(2)$ structure and Calabi-Yau metrics, whose harmonic forms can be
written down, and which yields exact topological information.

We were able to compute the metric on Calabi-Yau metric moduli space
and show that the restriction from the $\SU(2)$ structure moduli space
metric agrees with the Calabi-Yau moduli space metric computed in the
conventional way from the special geometry determined by classical
triple intersection numbers of $\CX_{m,n}$.  The light scalars of the
$\CN=4$ theory were determined by a twisted cohomology ring associated
with the $SU(2)$ structure, of which only the exact Calabi-Yau moduli
of the $\CN=2$ theory remain in the standard cohomology ring.  For
nonminimal topological data $(m,n)\ne(1,1)$, the lifted cohomology
classes of the twisted cohomology ring persist as torsion classes of
the standard cohomology ring.

Finally, to be able to provide precise statements in integer homology,
we have extended the results of Ref.~\cite{Donagi:2008ht} on the
integer homology ring of $\CX_{m,n}$.  This has allowed us to relate a
basis of cohomology classes from the first-order analysis of harmonic
forms to exact integer homology classes of~$\CX_{m,n}$ based on
embeddings of Hodge surfaces of $\CX_{m,n}$.

The dual $T^6/\IZ_2$ orientifold with $\CN=2$ flux is a simple model
of a type IIB warped compactification, embodying features of more
realistic models.  In future work, we plan to apply the results of
this investigation to provide a derivation of the procedure for warped
Kaluza-Klein reduction of type IIB string theory on $T^6/\IZ_2$ by
duality to standard compactification of type IIA string theory on the
class of Calabi-Yau manifolds $\CX_{m,n}$.  Nearly all string
compactifications of phenomenological interest are warped, yet
explicit examples of warped Kaluza-Klein reduction are almost
nonexistent, despite a promising proposal for a general framework in
which to understand them~\cite{Shiu:2008ry,Douglas:2008jx,
  Frey:2008xw,Frey:2009qb,Underwood:2010pm}, building on earlier
work~\cite{DeWolfe:2002nn,Giddings:2005ff} We hope to provide a
complementary approach to
Refs.~\cite{Shiu:2008ry,Douglas:2008jx,Frey:2008xw}, and to provide
explicit examples with which to probe that formalism.  

Another interesting line of research concerns connections between the
$\CX_{m,n}$ for different $(m,n)$.  By going to a point in complex
structure modulus at which $\MW(\CX_{1,1})$ develops $\IZ_m$ torsion,
it appears possible to resolve the singular manifold in such a way
that the new principally polarized abelian fibration has
$\pi_1=\IZ_m$.  Similar $\pi_1$ changing topological transitions have
been considered in Refs.~\cite{Davies:2009ub,Davies:2011is}.  The
transition relate $\CX_{1,1}$ to a new manifold that lies midway
between $\CX_{m,1}$ and $\CX_{1,m}$ in the sense that lifting to an
$m$-fold cover gives the former and quotienting by $\IZ_m$ gives the
latter.  These transitions are currently under
investigation~\cite{DonagiSchulz}.  The manifolds $\CX_{m,n}$ are of
potential interest in heterotic model building since they have very
few moduli and fundamental groups $\IZ_n\times\IZ_n$ useful for Wilson
lines.  It is worth understanding transitions of this sort, which
could yield other new manifolds of phenomenological interest.

%%%%%%%%%%%%%%%%%%%%%%%%%%%%
%%%   Acknowledgements   %%%
%%%%%%%%%%%%%%%%%%%%%%%%%%%%

\bigskip\centerline{\bf Acknowledgments}\nobreak\medskip\nobreak

I am grateful to A.-K. Kashani-Poor, R.~Donagi, and E.~Tammaro for
helpful conversations.  In addition, I would like to thank the KITP
Scholars program and the University of Pennsylvania for their
continued hospitality.  This material is based upon work supported by
the National Science Foundation under Grant No.~PHY09-12219.  This
research was supported in part by the National Science Foundation
under Grant No.~PHY11-25915.

%%%%%%%%%%%%%%%%%%%%%%
%%%   Appendices   %%%
%%%%%%%%%%%%%%%%%%%%%%
\appendix

%%%%%%%%%%%%%%%%%%%%%%%%%%%%%%%%%%%%%%%%%
%%%   A. Kahler moduli space metric   %%%
%%%%%%%%%%%%%%%%%%%%%%%%%%%%%%%%%%%%%%%%%

\section{K\"ahler moduli space metric}
\label{app:Kahler}

In this Appendix, we show that the K\"ahler
potential~\eqref{eq:KahlerPotv} indeed leads to the
metric~\eqref{eq:sKahlerMetric} on K\"ahler moduli space.  Let us
complexify the $v^A$ of Sec.~\ref{sec:KandCmetrics} by writing $t^A =
b^A+iv^A$.  Then, the K\"ahler potential~\eqref{eq:KahlerPotv} becomes
\begin{equation}
  \begin{split}
    K &= -\log 8V = -\log\frac{i}{3!}\frac{m}{n}C_{ABC}(t^A-t^A)(t^B-t^B)(t^C-t^C)\\
    &= -\log\frac{m}{n}(t^2-\tbar^2)\bigl((t^1-\tbar^1)(t^2-\tbar^2)-(\bt-\bar\bt)^2\bigr)\\
    &= -\log (t^2-\tbar^2)-\log\bigl((t^1-\tbar^1)(t^2-\tbar^2)-(\bt-\bar\bt)^2\bigr)
    + \text{const.}\vphantom{\frac{m}{n}}
\end{split}
\end{equation}
Differentiating gives K\"ahler metric $K_{A\bar B} = \pd_A\bar\pd_B
K$, with components
\begin{equation}
  \begin{split}
    4K_{1\bar1} &= \frac{(v^2)^2}{(v^1v^2-\bv^2)^2},\\
    4K_{2\bar2} &= \frac1{(v^2)^2} + \frac{(v^1)^2}{(v^1v^2-\bv^2)^2},\\
    4K_{1\bar2} &= -\frac1{v^1v^2-\bv^2} + \frac{v^1v^2}{(v^1v^2-\bv^2)^2},\\
    4K_{I\Jbar} &= \frac{2\d_{IJ}}{v^1v^2-\bv^2} + \frac{4v^Iv^J}{(v^1v^2-\bv^2)^2},\\
    4K_{1\Ibar} &=-\frac{2v^Iv^2}{(v^1v^2-\bv^2)^2},\\
    4K_{2\Ibar} &=-\frac{2v^Iv^1}{(v^1v^2-\bv^2)^2},
  \end{split}
\end{equation}
and $K_{B\bar A} = K_{A\bar B}$.
%\begin{subequations}
% \begin{align}
%    K_{1\bar1} &= \frac{(t^2-\tbar^2)^2}{((t^1-\tbar^1)(t^2-\tbar^2)-(\bt-\bar\bt))^2},\\
%    K_{2\bar2} &= \frac1{((t^2-\tbar^2))^2} 
%   + \frac{((t^1-\tbar^1))^2}{((t^1-\tbar^1)(t^2-\tbar^2)-(\bt-\bar\bt))^2},\\
%    K_{1\bar2} &= -\frac1{(t^1-\tbar^1)(t^2-\tbar^2)-(\bt-\bar\bt)} 
%    + \frac{(t^1-\tbar^1)(t^2-\tbar^2)}{((t^1-\tbar^1)(t^2-\tbar^2)-(\bt-\bar\bt))^2},\\
%    K_{I\Jbar} &= \frac{2\d_{IJ}}{(t^1-\tbar^1)(t^2-\tbar^2)-(\bt-\bar\bt)2} 
%    + \frac{4(t^I-\tbar^I)(t^J-\tbar^J)}{((t^1-\tbar^1)(t^2-\tbar^2)-(\bt-\bar\bt))^2},\\
%    K_{1\Ibar} &=-\frac{2(t^I-\tbar^I)(t^2-\tbar^2)}{((t^1-\tbar^1)(t^2-\tbar^2)-(\bt-\bar\bt))^2},\\
%    K_{2\Ibar} &=-\frac{2(t^I-\tbar^I)(t^1-\tbar^1)}{((t^1-\tbar^1)(t^2-\tbar^2)-(\bt-\bar\bt))^2}.
%  \end{align}
%\end{subequations}
%
Then, restricting to $t^A=iv^A$ purely imaginary, we find
\begin{equation}
  \begin{split}
    4ds^2_\text{K\"ahler} &= 4K_{A\bar B}\d v^A \d v^B\\
    &= \Bigl(\frac{\d v^2}{v^2}\Bigr)^2
    +\frac{1}{(v^1v^2-\bv^2)^2}\bigl(\d(v^1v^2-\bv^2)\bigr)^2
    -\frac{2}{v^1v^2-\bv^2}\bigl(\d v^1\d v^2 -(\d\bv)^2\bigr).
  \end{split}
\end{equation}
If we introduce new variables $s_1,s_2,x^{3I}$ via
\begin{equation}
  \begin{split}
    v^2 &= s_2,\\
    v^1 &= s_1 - s_2\d_{IJ}x^{3I}x^{3J},\\
    v^I &= s_2x^{3I},
  \end{split}
\end{equation}  
then $v^1v^2-\bv^2 = s_1s_2$, and it is possible to show that
\begin{equation}
  (\d\bv)^2 - \d v^1\d v^2 = (s_2)^2(\d\bx^3)^2 - \d s_1\d s_2.
\end{equation}
Therefore, the previous expression for the K\"ahler metric becomes
\begin{equation}
  \begin{split}
    4ds^2_\text{K\"ahler}
    &= \Bigl(\frac{\d s_2}{s_2}\Bigr)^2
    +\Bigl(\frac{\d(s_1s_2)}{s_1s_2}\Bigr)^2
    -\frac{2}{s_1s_2}\big(s_2(\d\bx^3)^2 -\d s_1\d s_2\bigr)\\
    &= 2\Bigl(\frac{\d s_2}{s_2}\Bigr)^2
    +\Bigl(\frac{\d s_1}{s_1}\Bigr)^2
    -\frac{2}{s_1s_2}\big(s_2(\d\bx^3)^2\bigr),\\
  \end{split}
\end{equation}
as claimed in Sec.~\ref{sec:KandCmetrics}.

%%%%%%%%%%%%%%%%%%%%%%%%%%%%%%%%%%%%%%%%%%%%%%%%%%%%%%%%%
%%%   B. Intersections of Hodge surfaces in X_{m,n}   %%%
%%%%%%%%%%%%%%%%%%%%%%%%%%%%%%%%%%%%%%%%%%%%%%%%%%%%%%%%%

\section{Intersections of Hodge surfaces in $\CX_{m,n}$}
\label{app:Intersections}
 
In this Appendix, we compute the double and triple intersections of
integer divisors in the Calabi-Yau manifolds $\CX_{m,n}$, extending
the results of Ref.~\cite{Donagi:2008ht}.

For the case $(m,n)=(1,1)$, the Calabi-Yau manifold $\CX_{1,1}$ was
realized in the second construction of Ref.~\cite{Donagi:2008ht} as
the relative Jacobian $\Pic^0(S/\IP^1)$ of a surface $S$ fibered by
genus-2 curves over $\IP^1$.  In this case, the Hodge curves and
surfaces of Secs.~\ref{sec:HodgeSurfaces} are known as theta curves
and theta surfaces.

As explained Sec.~4.3.3 and App.~J of Ref.~\cite{Donagi:2008ht}, the
sections of $S$ can be viewed as sections of $\Pic^1(S/\IP^1)$, while
those of $\CX_{1,1}$ can be viewed as sections of $\Pic^0(S/\IP^1)$.
To relate the two, we note that $\Pic^0(S/\IP^1)\cong \Pic^0(S/\IP^1)$
under a noncanonical isomorphism obtained by tensoring with a
privileged section of $S$,
\begin{equation}\label{eq:Picisomorph}
  \Pic^1(S/\IP^1)
  \quad \xrightarrow{\otimes [\ell_0]^{-1}}\quad
  \Pic^0(S/\IP^1),
  \qquad 
  \ell \mapsto \ell-\ell_0,
\end{equation}
where $\ell_0\in\{\ell_I,\ell'_I\}$.  The isomorphism depends on the
choice of which of the $2M=24$ sections $\ell_I,\ell'_I$ of $S$ maps
to the zero section of $\CX$.

Similarly, each section of $\Pic^1(S/\IP^1)$ determines a theta
surface, embedding $S$ into $\Pic^0(S/\IP^1)$.  Thus, for the $2M$
choices $\ell_0=\ell_I$ and $\ell'_I$, for $I=1,\dots,M$, we obtain
$2M$ theta surfaces $\Theta_I$ and $\Theta'_I$ in
$\CX_{1,1}\cong\Pic^0(S/\IP^1)$.  App.~J of Ref.~\cite{Donagi:2008ht}
computed the intersections of these theta surfaces to be
\begin{equation}\label{eq:DblInts}
  \begin{split}
    \Theta_I\cdot\Theta_J &= \s_0 + \s_{\ell'_I-\ell_J},\quad
    \Theta'_I\cdot\Theta'_J = \s_0 + \s_{\ell_I-\ell'_J},\\
    \Theta'_I\cdot\Theta_J &= \s_0 + \s_{\ell_I-\ell_J},\quad
    \Theta_I\cdot\Theta_I' = 2\s_0 + C_{\ell_I\cap\ell'_I}.
  \end{split}
\end{equation}
Here, $L\to \s_L$ is the isomorphism identifying degree zero line
bundles in $\Pic^0(S/\IP^1)$ with sections of the abelian fibered
threefold $\CX$. The curve $C_{\ell_I\cap\ell'_I}$ is the common
genus-2 fiber of $\Theta_I$ and $\Theta'_I$.  Note that
$\ell'_I-\ell_J = \ell'_J-\ell_I$ as a consequence of
\begin{equation}
  [\ell_I+\ell'_I]=[C'],
  \quad\text{independent of $I$.}
\end{equation}
For self intersections, we have
\begin{equation}
  [\Theta_I\cdot\Theta_I] = c_1(K_{\Theta_I}),
\end{equation}
where for the surface $S$, it was shown that
\begin{equation}
  c_1(K_S) = [C'] - [C].
\end{equation}
Here, $[C]$ is the class of the genus-2 fiber of $S$.
The triple intersections were shown to be
\begin{equation}
  \begin{split}
    &\Theta_I\cdot\Theta_J\cdot\Theta_K =1,\\
    &\Theta_I\cdot\Theta_J\cdot\Theta_J=\Theta_I\cdot\Theta'_J\cdot\Theta'_J =-2,\\
    &\Theta_I\cdot\Theta_I\cdot\Theta'_I=\Theta_I\cdot\Theta_J\cdot\Theta'_J =0,\\
    &\Theta_I\cdot\Theta_I\cdot\Theta_I=-4,
  \end{split}
  \quad
  \begin{split}
    & A\cdot\Theta_I\cdot\Theta_J = A\cdot\Theta_I\cdot\Theta'_J = 2,\\
    & A\cdot\Theta_I\cdot\Theta_J =2,\\
    &A\cdot\Theta_I\cdot\Theta_J =2,\\
    &A\cdot\Theta_I\cdot\Theta_I=2.
  \end{split}
\end{equation}
together with equations obtained from these by exchange of $\Theta$ and
$\Theta'$.

%%%%%%%%%%%%%%%%%%%%%%%%%%%%%%%%%%%%%%%%%%
%%%   Double intersections in X_{m,n}  %%%
%%%%%%%%%%%%%%%%%%%%%%%%%%%%%%%%%%%%%%%%%%

\subsubsection*{Double intersections in $\CX_{m,n}$}

For general $(m,n)$ we claim that the generalization of
Eq.~\eqref{eq:DblInts} for intersections in $\CX_{m,n}$ is
\begin{equation}\label{eq:DblIntsGeneral}
  \begin{split}
    [S_I\cdot S_J] &= \mbar\nbar[\s_0] + \mbar\nbar[\s_{\ell'_I-\ell_J}],\quad
    [S'_I\cdot S'_J] = \mbar\nbar[\s_0] +\mbar\nbar[\s_{\ell_I-\ell'_J}],\\
    [S'_I\cdot S_J] &= \mbar\nbar[\s_0] + \mbar\nbar[\s_{\ell_I-\ell_J}],\quad
    [S_I\cdot S_I'] = 2\mbar\nbar[\s_0] + \mbar^2\nbar[C_{\ell_I\cap\ell'_I}],
  \end{split}
\end{equation}
with self intersections given by
\begin{equation}\label{eq:SelfIntGeneral}
  [S_I\cdot S_I] = c_1(K_{S_I}),
  \quad\text{with}\quad K_S = \mbar\nbar [C'] - \mbar^2\nbar[C].
\end{equation}
That two Hodge surfaces generically intersect in $2\mbar\nbar$
sections is clear fiberwise, since in each $T^4$ fiber $A$, two
genus-$(1+\mbar\nbar)$ curves $C$ generically intersect $2\mbar\nbar$
points (cf.~Eq.~\eqref{eq:mnEuler}).  On the other hand the subscripts
in Eq.~\eqref{eq:DblIntsGeneral} require explanation.  Since $S$ is
fibered by genus $\mbar\nbar$ curves and $\Pic^0{S/\IP^1}$ is a
$T^{2+2\mbar\nbar}$ fibration over $\IP^1$, it is not immediately
clear why sections of $\CX_{m,n}$ should be associated with sections
of $\Pic^0(S/\IP^1)$.  Let us focus on the case $(m,n)=(1,n)$.  In
this case, we have a map $f\colon\Pic^0(S/\IP^1)\to\CX_{1,n}$ acting
fiberwise as
\begin{equation}
  (x^1,x^2;y^{1i},y^{2i})\mapsto (x^1,x^2,y^1,y^2) = 
  (x^1,x^2;\sum_{i=1}^n y^{1i},\sum_{i-1}^n y^{2i}).
\end{equation}
The inverse map $f^{-1}$ gives an embedding of an $n^2$-fold cover of
$\CX_{1,n}$ into $\Pic^0(S/\IP^1)$,
\begin{equation}
 (x^1,x^2;y^1,y^2)\mapsto(x^1,x^2,y^{1i},y^{2i}),
 \quad\text{with}\quad y^{1i}=y^1/n,\ y^{2i} = y^2/n.
\end{equation}
This map takes the principally polarized $T^{2+2n}$ fibration
$\Pic^0(S/\IP^1)$ to the $(n,1)$ polarized $T^4$ fibration $\CX_{1,2}$.

Let us show that the polarizations are as claimed.  Under the inverse
map, the K\"ahler form
\begin{equation}
  J_\text{fib} = h(dx^1\w dx^2 + \sum_{i=1}^n dy^{1i}\w dy^{2i})
\end{equation}
on the $T^{2+n}$ fibers of $\Pic^0(S/\IP^1)$ pulls back to K\"ahler form
\begin{equation}
  (f^{-1})^*J_\text{fib} 
  = h(dx^1\w dx^2 + \sum_{i=1}^n \frac{dy^1}{n}\w\frac{dy^2}{n})
  = h' (n dx^1\w dx^2 + dy^1\w dy^2)
\end{equation}
on the $T^4$ fibers of $\CX_{1,n}$, where $h'=h/n$.  Alternatively,
since $\CX_{1,n} = \CX_{n,1}/(\IZ_n\times\IZ_n$), we can view $(\hat
y^1,\hat y^2) = (y^1/n,y^2/n)$ as coordinates on the fibers of
$\CX_{n,1}$, and the previous pullback gives fiber K\"ahler form
\begin{equation}
   h (dx^1\w dx^2 + n d\yhat^1\w d\yhat^2)
\end{equation}
on the fibers of $\CX_{n,1}$.  The remaining case $(m,n)=(2,2)$ is
obtained as $\CX_{2,2} = \CX_{4,1}/(\IZ_2\times\IZ_2)$.

%%%%%%%%%%%%%%%%%%%%%%%%%%%%%%%%%%%%%%%%%%%
%%%   Triple intersections in X_{m,n}   %%%
%%%%%%%%%%%%%%%%%%%%%%%%%%%%%%%%%%%%%%%%%%%

\subsubsection*{Triple intersections in $\CX_{m,n}$}

The triple intersections can be obtained as double intersections of
curves in surfaces.  For example, for $I,J,K$ distinct,
\begin{equation}\label{eq:TrpIntIJK}
  \begin{split}
    S_I\cdot S_J\cdot S_K &= 
    (S_I\cdot S_J)\cdot_{S_J}(S_J\cdot S_K)\\ 
    &= (\mbar\nbar[\s_0]+\mbar\nbar[\s_{\ell'_I-\ell_J}])
    \cdot_{S_J}(\mbar\nbar[\s_0]+\mbar\nbar[\s_{\ell'_K-\ell_J}])\\ 
    &\cong \mbar^2\nbar^2 (\ell_J+\ell'_I)\cdot_S(\ell_J+\ell'_K)\\ 
    &= -\mbar^3\nbar^2.
  \end{split}
\end{equation}
Here, we have used the fact that $\s_{\ell-\ell_J}$ maps to $\ell\in S$ under
the isomorphism $S_J\to S$.   The remaining triple intersections of theta
surfaces are
\begin{equation}\label{eq:TrpIntList}
  \begin{split}
    & S_I\cdot S_J\cdot S_J= S_I\cdot S'_J\cdot S'_J =-2\mbar^3\nbar^2,\\
    & S_I\cdot S_I\cdot S'_I= S_I\cdot S_J\cdot S'_J =0,\\
    & S_I\cdot S_I\cdot S_I=-4\mbar^3\nbar^2,\\
  \end{split}
\end{equation}
together with equations obtained from these by exchange of $S$ and $
S'$.  The computation is analogous to the previous one.  

From the independence the result on the choice of which of the three
Hodge surfaces is used to perform the double intersection, we can
deduce the coefficients of $[C]$ and $[C']$ in the double intersection
expressions~\eqref{eq:DblIntsGeneral} and \eqref{eq:SelfIntGeneral}.
For example, supposed that
\begin{equation}
  [S_I\cdot S'_I] = 2\mbar\nbar[\s_0] + \lambda
  [C_{\ell_I\cap\ell'_I}].
\end{equation}
Then, for agreement of
\begin{equation}
  \begin{split}
    S_I\cdot S'_I\cdot S_J &= 
    (S_I\cdot S'_I)\cdot_{S_I}(S_I\cdot S_J)\\ 
    &= (2\mbar\nbar[\s_0]+\lambda [C_{\ell_I\cap\ell'_I}])
    \cdot_{S_I}(\mbar\nbar[\s_0]+\mbar\nbar[\s_{\ell'_J-\ell_I}])\\ 
    &\cong \mbar^2\nbar^2 (2\ell_I+\lambda[C])\cdot_S(\ell_I+\ell'_J)\\ 
    &= -2\mbar^3\nbar^2 + 2\lambda\mbar\nbar,
  \end{split}
\end{equation}
and
\begin{equation}
  \begin{split}
    S_I\cdot S'_I\cdot S_J &= 
    (S_I\cdot S_J)\cdot_{S_J}(S'_I\cdot S_J)\\ 
    &= (\mbar\nbar[\s_0]+\mbar\nbar[\s_{\ell'_I-\ell_J}])
    \cdot_{S_J}(\mbar\nbar[\s_0]+\mbar\nbar[\s_{\ell_I-\ell_J}])\\ 
    &\cong \mbar^2\nbar^2 (\ell_J+\ell'_I)\cdot_S(\ell_J+\ell_I)\\ 
    &= 0,
  \end{split}
\end{equation}  
we require $\lambda = \mbar^2\nbar$.
Similarly, writing $K_S = \a[C']-\b[C]$, we find from
\begin{equation}
  S_I\cdot S_I\cdot S_J = (S_I\cdot S_I)\cdot_{S_I}(S_I\cdot S_J)
  = (S_I\cdot S_J)\cdot_{S_J}(S_I\cdot S_J)
\end{equation}
that $\b=\mbar^2\nbar$, and then from
\begin{equation}
  S_I\cdot S_I\cdot S'_I = (S_I\cdot S_I)\cdot_{S_I}(S_I\cdot S'_I)
  = (S_I\cdot S'_I)\cdot_{S'_I}(S_I\cdot S'_I)
\end{equation}
that $\a = \mbar\nbar$.

Finally, let us compute the triple intersections involving the generic
abelian surface fiber~$A$. In this case, $A^2=0$, and
\begin{equation}
  A\cdot S_I\cdot S_J = A\cdot S_I\cdot S'_J 
  = A\cdot S'_I\cdot S'_J = 2\mbar\nbar,
\end{equation}
for any $I,J$, not necessarily distinct.  This is most easily proven from the
intersection of curves in the abelian fiber $A$.  For example,
\begin{equation}
  \begin{split}
    A\cdot S_I\cdot S_J
    &= (A\cdot S_I)\cdot_A(A\cdot S_J)\\ 
    &\cong C\cdot_A C=2\mbar\nbar,
  \end{split}
\end{equation}
as desired.  (In an abelian surface, the self-intersection of a
genus-$g$ curve is $2g-2$, and we have $g=1+\mbar\nbar$.)

The same result is obtained if the intersections are performed in a theta
surface.  Let $C_I$ denote the genus-2 fiber of $ S_I\cong S$.  Then, for
example, for $I\ne J$,
\begin{equation}
  \begin{split}
    A\cdot S_I\cdot S_J
    &= (A\cdot S_I)\cdot_{ S_I}( S_I\cdot S_J)\\
    &= C_I\cdot_{S_I}(\mbar\nbar[\s_0]+\mbar\nbar[\s_{\ell'_J-\ell_I}])\\
    &\cong C\cdot_S(\mbar\nbar[\ell_I]+\mbar\nbar[\ell'_J])=2\mbar\nbar,\\
    A\cdot S_I\cdot S_I
    &= (A\cdot S_I)\cdot_{S_I}[S_I\cdot S_I])\\
    &= C_I\cdot_{ S_I} [c_1(K_{S_I}])\\
    &\cong [C]\cdot_S [\mbar\nbar C'-\mbar^2\nbar C)]=2\mbar\nbar.
  \end{split}
\end{equation}
In the last step, we have used the fact that the genus
$g=1+\mbar\nbar$ fiber $C$ of $S$ satisfies $C^2=0$ and $C\cdot C' =
C\cdot(\ell_K+\ell'_K) = 2$, when the intersections are computed in
$S$.

%%%%%%%%%%%%%%%%%%%%%%
%%%   References   %%%
%%%%%%%%%%%%%%%%%%%%%%

\end{document}